%% file: main.tex
\documentclass[11pt]{article}
\usepackage[letterpaper,margin=1in]{geometry}
\usepackage[T1]{fontenc}
\usepackage{lmodern}
\usepackage{microtype}
\usepackage{graphicx}
\usepackage{amsmath,amssymb,amsthm,mathtools}
\usepackage{enumitem}
\usepackage{booktabs,tabularx,array,multirow}
\usepackage[hidelinks]{hyperref}
\usepackage{xcolor}
\usepackage{algorithm}
\usepackage[noend]{algorithmic}
\usepackage{tikz}
\usetikzlibrary{arrows.meta,decorations.pathreplacing}
\usetikzlibrary{positioning,arrows,fit}
\usetikzlibrary{matrix}
\newcommand{\figurefontsize}{\fontsize{11}{13.2}\selectfont}

\setlist[itemize]{topsep=0.2em,itemsep=0.2em,leftmargin=1.4em}
\setlist[enumerate]{topsep=0.2em,itemsep=0.2em,leftmargin=1.8em}
\allowdisplaybreaks
\renewenvironment{abstract}{\begin{center}\bfseries Abstract\end{center}\noindent}{\par}
\hypersetup{
  pdftitle={Linear Set Reconciliation with O(d) Additive Redundancy},
  pdfauthor={Tianren Liu, Rui Xu, Tong Yang, Kangyang Zhou},
  pdfsubject={Set reconciliation and binary linear sparse recovery}
}

\newtheorem{theorem}{Theorem}[section]
\newtheorem{lemma}[theorem]{Lemma}
\newtheorem{proposition}[theorem]{Proposition}
\newtheorem{corollary}[theorem]{Corollary}

\newtheorem{remark}[theorem]{Remark}
\newtheorem{definition}[theorem]{Definition}
\newtheorem{problem}[theorem]{Problem}

\newcommand{\eps}{\varepsilon}
\newcommand{\wt}[1]{\left\lVert #1 \right\rVert_0}
\newcommand{\supp}{\operatorname{supp}}

\newcommand{\ceil}[1]{\left\lceil #1 \right\rceil}

\newcommand{\polylog}{\operatorname{polylog}}
\newcommand{\poly}{\operatorname{poly}}

\newcommand{\F}{\mathbb{F}}

\newcommand{\HintSketch}{\HSketch}

\newcommand{\bgrev}{\color{blue}}
\newcommand{\edrev}{\color{black}}

\newcommand{\HSketch}{\textsc{HintSketch}}

\newcommand{\CR}{\mathsf{CR}}

\input{2609-version/ai-revision-format}

\begin{document}

\begin{titlepage}
\title{\edrev Linear Set Reconciliation with $O(d)$ Additive Redundancy\edrev}
\author{\edrev\begin{tabular}{cc}
Tianren Liu & Rui Xu \\
Peking University & Peking University \\[1em]
Tong Yang & Kangyang Zhou \\
Peking University & Massachusetts Institute of Technology
\end{tabular}\edrev}
\date{}
\maketitle
\thispagestyle{empty}
\input{2609-version/0abstract}
\end{titlepage}
\input{2609-version/1intro}
\input{2609-version/2preliminaries}

\newpage \input{2609-version/3construction}
\input{2609-version/4prob-tools}
\input{2609-version/5hintsketch}
\input{2609-version/6toplevel}
\bgrev
\edrev
\nocite{CL20}

\input{2609-version/Z-AIdisclosure}

\clearpage
\bibliographystyle{alpha}
\bibliography{2609-version/ref}

\clearpage
\appendix

\input{2609-version/B-probabilistic-tools}

\end{document}

%% file: 2609-version/0abstract.tex
\begin{abstract}
In set reconciliation, two parties holding sets $A, B \subseteq [n]$ with $|A \triangle B| \le d$ seek to recover their symmetric difference from short $\F_2$-linear sketches, which is equivalent to recovering a $d$-sparse vector $x = \mathbf{1}_{A \triangle B} \in \F_2^n$ from its syndrome $Hx \in \F_2^m$.
While the information-theoretic counting bound requires $L(n,d) := \log \sum_{i=0}^d \binom{n}{i} = d\log(n/d) + \Theta(d)$ bits, existing polynomial-time schemes either use at least $d\ceil{\log(n+1)}$ bits (algebraic sketches, leaving an $\Omega(d\log d)$-bit gap) or incur a multiplicative $(1+\delta)L(n,d)$ overhead (peeling-based sketches), raising the question of whether $L(n,d) + O(d)$ bits is achievable with fast updates and decoding.

We construct a public-coin randomized $\F_2$-linear sketch of length $L(n,d) + O(d) + \lambda\log^{O(1)} n$ that recovers each fixed input of weight at most $d$ with probability at least $1 - 2^{-\lambda}$, with $\log^{O(1)} n$ update time and $d\exp(O((\log\log n)^2)) = dn^{o(1)}$ worst-case decoding time.
In particular, for $n \le 2^{\polylog(d)}$ and $\lambda \le d^{1-\Omega(1)}$, this achieves $L(n,d) + O(d)$ bits with $\polylog(d)$ update time and $d^{1+o(1)}$ decoding time.

Our construction encodes $d$ permuted rows via randomly scaled Pinsketch syndromes and sketches the resulting bit array column by column. Decoding columns sequentially from prefix predictions cancels recovered rows and exposes undecided rows as decoder-side \emph{hints}, reducing each column to a \emph{hinted sparse recovery} problem that we solve via a recursive sketch, $\HintSketch$.
\end{abstract}

%% file: 2609-version/1intro.tex
\section{Introduction}
\label{sec:intro}

In the \emph{set reconciliation} problem, two parties hold sets $A,B\subseteq[n]$ with $|A\triangle B|\le d$ and seek to recover their symmetric difference while communicating as little information as possible.
The problem is important in distributed data management~\cite{MTZ03,GM11,EGUV11}, blockchain communication~\cite{Graphene19,YGA24}, and network monitoring~\cite{LossRadar16,ChameleMon23}.
Moreover, the same problem arises in error-correcting codes and sparse recovery over finite fields~\cite{DORS08,CR23}.

Identifying sets with their characteristic vectors $\mathbf{1}_A,\mathbf{1}_B\in \F_2^n$, a binary linear sketch $H \in \F_2^{m \times n}$ satisfies $H\mathbf{1}_A + H\mathbf{1}_B = H\mathbf{1}_{A \triangle B} \pmod 2$, reducing set reconciliation to recovering a vector $x \in \F_2^n$ of weight $\wt{x} \le d$ from its \emph{syndrome} $Hx \in \F_2^m$.
Toggling an element $i \in [n]$ adds column $He_i$ to the sketch, so the maximum column weight $\max_i \wt{He_i}$ (\emph{column locality}) governs per-update time.

Information-theoretically, any randomized sketch that recovers each fixed input of weight at most $d$ with probability at least $1-\eps$ requires at least $L(n,d) + \log(1-\eps)$ bits, where
\[
L(n,d) := \log \sum_{i=0}^d \binom{n}{i} = d\log(n/d) + \Theta(d) \qquad (\text{for } d \le n/2).
\]
Existing schemes exhibit a trade-off between sketch length and computational efficiency.

\textbf{1)} Classical algebraic schemes represented by characteristic polynomials~\cite{MTZ03} and BCH-based Pinsketch~\cite{DORS08} store $\approx d\log n$ bits, which exceeds $L(n,d)$ by $\Theta(d\log d)$ bits (e.g., $2d\log d$ bits vs.\ $d\log d + O(d)$ bits when $n = d^2$) and requires $\Omega(d)$ update time.

\textbf{2)} Hashing and peeling sketches such as Invertible Bloom Lookup Tables (IBLTs) and their recent spatial-coupling refinements~\cite{GM11,HPW23,BKW24,FLOS24,Wal25,XYZ26,KPW26} support fast updates and nearly linear decoding.
For instance, Stuffed IBLTs~\cite{KPW26} can approach the lower bound within any constant factor $1+\delta$.
However, this still incurs $\Theta(\delta d\log(n/d))$ redundant bits beyond $L(n,d)$, which is $\omega(d)$.

\textbf{3)} A straightforward random linear sketch can indeed solve this problem in $L(n,d)+\log(1/\eps)+O(1)$ bits.
However, no polynomial-time decoder is known for this random sketch~\cite{FA25}, making it less suitable for practical usage where $n,d$ can be large.
In fact, recovering the sparse input from its syndrome is the syndrome decoding problem, which is even conjectured to be average-case hard in cryptography~\cite{Ste94}.

This raises a natural question:
\begin{quote}
\emph{Can a linear sketch achieve length $L(n,d) + O(d)$ --- only $O(d)$ bits of additive redundancy above the counting bound --- while retaining fast updates and decoding?}
\end{quote}

\subsection{Our Results}
We answer this question affirmatively in the public-coin model, where randomness (including random permutations, random coefficients, \textit{etc.}) is shared. Put $\lambda = \ceil{\log(1/\eps)}$.

\begin{theorem}[Main result: Theorem~\ref{thm:complete-main}]
\label{thm:intro-main}
For $2 \le d \le n$ and $0 < \eps \le 1/2$, there is a public-coin randomized $\F_2$-linear sketch $\mathcal H_r$ of length
\[
L(n,d) + O(d) + \lambda \log^{O(1)} n
\]
together with a decoder \(\mathsf{Dec}_r\) satisfying
$$
\forall x\in\mathbb F_2^n\ \text{with }\wt{x}\le d,
\qquad
\Pr_r\!\left[\mathsf{Dec}_r(\mathcal H_r x)=x\right]\ge1-\varepsilon,
$$
where $r$ is the randomness used in the sketch.
Our algorithm also achieves:
\begin{itemize}
    \item \textbf{Polylogarithmic updates:} the time complexity of each update is $\log^{O(1)} n$.
    \item \textbf{Fast decoding:} the worst-case decoding time is $d\exp(O((\log\log n)^2)) = d n^{o(1)}$.
\end{itemize}
\end{theorem}

\begin{corollary}[$O(d)$ additive redundancy: Corollary~\ref{cor:near-counting}]
\label{cor:intro-near-counting}
For fixed constants $A_0, c_0 > 0$, if $n \le 2^{(\log d)^{A_0}}$ and $\lambda \le d^{1-c_0}$, the sketch uses $L(n,d) + O(d)$ bits, with $\polylog d$ update time, and $d^{1+o(1)}$ worst-case decoding time.
\end{corollary}

Table~\ref{tab:comparison} compares representative space and time bounds. 

\begin{table}[!htbp]
\edrev
\begin{center}
\renewcommand{\arraystretch}{1.1}
\setlength{\tabcolsep}{5pt}
\begin{tabular}{c|c|c|c}
\specialrule{2pt}{0em}{0em}
\textbf{Scheme} & \textbf{Sketch length (bits)} & \textbf{Update} & \textbf{Decoding} \\
\specialrule{1pt}{0em}{0em}
Pinsketch & \multirow{2}{*}{$d\ceil{\log(n+1)}$} & \multirow{2}{*}{$O(d)$} & \multirow{2}{*}{$\widetilde O(d^2)$} \\
\cite{DORS08,Str12} & & & \\
\hline
CR sketch & \multirow{2}{*}{$O(d\log^2 d)$} & \multirow{2}{*}{$O(\log^2 d)$} & \multirow{2}{*}{$O(d\log^2 d)$} \\
\cite{CR23} & & & \\
\hline
Classical IBLT & \multirow{2}{*}{$O(d\log d)$} & \multirow{2}{*}{$O(1)$} & \multirow{2}{*}{$O(d)$} \\
\cite{GM11} & & & \\
\hline
Stuffed IBLTs & \multirow{2}{*}{$(1+\delta)L(n,d)+O(d)$} & \multirow{2}{*}{$O_\delta(1)$} & \multirow{2}{*}{$O_\delta(d)$ expected} \\
\cite{KPW26} & & & \\
\hline
\multirow{2}{*}{Random linear sketch} & \multirow{2}{*}{$L(n,d)+O(\log d)$} & \multirow{2}{*}{$O(L(n,d))$} & \multirow{2}{*}{$\gg\poly(d)$\textsuperscript{$\dagger$}}\\
& & & \\
\hline
\multirow{2}{*}{\textbf{This work}} & \multirow{2}{*}{$L(n,d)+O(d)$} & \multirow{2}{*}{$\log^{O(1)} d$} & \multirow{2}{*}{$d^{1+o(1)}$} \\
& & & \\
\specialrule{2pt}{0em}{0em}
\end{tabular}
\end{center}
\caption{Representative bounds. We assume $n=\poly(d)$ and $\eps = 1/\poly(d)$ failure probability. Pinsketch and CR are deterministic. The IBLT rows use parameters for inverse-polynomial failure. Stuffed IBLTs assume $n\ge d^{1+\Omega(1)}$ and include hash storage. $\dagger$ Under the search syndrome decoding assumption for $n=d^{1+\Omega(1)}$~\cite{Ste94}.}
\label{tab:comparison}
\end{table} 

\subsection{Technical Overview}
\label{sec:intro-overview}

To achieve $L(n,d)+O(d)$ bits, we (1) encode rows independently and decode their columns in order; (2) use recursive $\HintSketch$ to recover each column from partial predictions; and (3) bound the total space needed for hints and incorrect predictions at the top level.

\subsubsection{Row bucketing and sequential column prediction}
Permuting $[n]$ into $m:=d$ balanced rows of length $N = \ceil{n/m}$ reduces each element's local row address to $w = \ceil{\log(N+1)} = \log(n/d) + O(1)$ bits (one ``word'').
If row $j$ contains $t_j$ differences ($\sum_{j=1}^d t_j \le d$), the first $t_j$ Pinsketch words ($t_j w$ bits) suffice to uniquely recover row $j$. If the row loads $t_j$ were known in advance, storing $t_j$ words for row $j$ would cost $\sum_{j=1}^d t_j w \le dw = L(n,d) + O(d)$ bits.
However, a linear sketch must be fixed \emph{before} the input is known, and random row loads vary up to $\Theta(\log d / \log\log d)$: allocating enough words for the maximum load wastes a superconstant factor, while allocating $O(1)$ words leaves a constant fraction of rows undecoded.

To adapt to unknown row loads with a fixed sketch, we \emph{encode horizontally along rows, but sketch and decode vertically across columns} (Figure~\ref{fig:intro-architecture}). Each row $x_j$ is encoded into an $\ell$-bit codeword $\mathsf{Enc}_j(x_j) \in \F_2^\ell$, and the encoder stores a sketch\footnote{$(H_s)_{s=1}^\ell$ are $\ell$ independent sketches whose detailed implementation will be discussed later.} $H_s C_s(x)$ of each column $C_s(x) \in \F_2^m$, plus a CR sketch $\CR_{n,t}x$~\cite{CR23} for residual recovery.

\input{2609-version/fig-intro-architecture}

We then decode the columns in order, from $s = 1$ to $\ell$. Before column $s$, each row decoder (\emph{worker}) $j$ uses its recovered $(s-1)$-bit prefix to predict its $s$-th bit $q_{j,s} \in \{0, 1, \perp\}$ (or $\perp$ if undecided). Let $J_s = \{j : q_{j,s} = \perp\}$ be the rows returning $\perp$ and $P_s \in \F_2^m$ be the prediction vector (with $0$ on $J_s$). By $\F_2$-linearity, the central coordinator (\emph{manager}) calculates the syndrome of the \emph{prediction residual} $D_s := C_s(x) + P_s$ from $H_s C_s(x)$ in the stored sketch and $P_s$:
\[
H_s D_s = H_s C_s(x) + H_s P_s.
\]
Once row $j$ has accumulated $t_j$ words, worker $j$ knows $x_j$ and predicts all subsequent bits correctly, so its entries in $D_s$ cancel to $0$ and consume zero capacity in all later columns.
Meanwhile, the remaining nonzero positions of $D_s$ split into two categories:
\begin{itemize}
\item \textbf{Known abstentions ($J_s$):} On $J_s$, the manager \emph{already knows which rows returned $\perp$} and only lacks their bit values, costing $\approx 1$ bit per coordinate.
If it is guaranteed that $|J_s|\le X_s$, this costs $\approx X_s$ bits total.
\item \textbf{Unknown prediction errors ($\supp(D_s) \setminus J_s$):}
Outside $J_s$, $D_s$ is nonzero only where a worker predicted the wrong bit.
If it is guaranteed that $|\supp(D_s) \setminus J_s|\le Y_s$, locating these $\le Y_s$ errors at unknown coordinates costs $O(h_m(Y_s))$ bits, where $h_m(Y) := Y\log(em/Y)$.
\end{itemize}
This reduces each column step to the following subproblem, with parameter $(N,X,Y)$:

\begin{problem}[Hinted sparse recovery; Problem~\ref{prob:hinted-recovery}]
\label{prob:intro-hinted}
Given a syndrome $Hx$ of an unknown vector $x \in \F_2^N$ and a decoder-side hint $J \subseteq [N]$ satisfying $|J| \le X$ and $|\supp(x) \setminus J| \le Y$, the goal is to recover $x$.
Note that while the parameter $(N,X,Y)$ and the randomness are public to both parties, the hint $J$ is known only to the decoder, not to the encoder.
\end{problem}

\subsubsection{Recursive {\mdseries\HintSketch} and charge contraction}
In this part, we will construct our \emph{hinted recovery sketch} ($\HintSketch$), solving the above problem with leading space term $X + \alpha h_N(Y)$, where $\alpha$ is a sufficiently large constant.

To obtain this result, $\HintSketch$ applies the same row-column strategy recursively, permuting $[N]$ into $m$ rows of length $\polylog(n)$ (with $O(1)$ expected outside-hint errors and $\polylog(n)$ hints per row) encoded by random binary sketch.
The row decoder enumerates only the outside-hint positions and solves a binary linear system over $\F_2$ for the hinted bits.
As the number of outside-hint positions in each row is small, the enumeration can be done in $\exp(O((\log\log n)^2))$ time.
For every column, we just use $\HintSketch$ recursively.

Through a series of careful technical estimates, we show that the space term does not blow up across recursive levels.
Specifically, a row with $|J_j|$ hinted positions and $r$ outside-hint positions can be specified using about $\sigma_{j,r}\approx |J_j|+r\log(eN_{\max})$ bits, where $N_{\max}=\ceil{N/m}$ is the maximum row length.
The worker tests a candidate with $r$ outside-hint positions only after collecting $\sigma_{j,r}+c$ bits, where $c=2\ceil{\log(N/Y)}+\Theta(1)$ is an extra wait.
This rule achieves two goals simultaneously:
\begin{enumerate}
\item \textbf{Unit-coefficient passthrough for hint cost $X$:} We prove that each row returns $\perp$ at most $|J_j|+y_j\log(eN_{\max})+c+O(1)$ times in expectation, where \(y_j\) is the actual number of nonzeros outside \(J_j\).
By adding small capacity margins, this gives $\sum_s X_s \le X + O( h_N(Y))$ hints to the children, preserving coefficient $1$ on $X$.
\item \textbf{Geometric contraction of outside-hint cost $Y$ and density $\rho = Y/N$:} Waiting $c$ extra parity bits suppresses false candidates by $2^{-c} \le 2^{-\Theta(1)}(Y/N)^2$, shrinking both the children's outside-hint location cost and the \emph{density} defined as $\rho := Y/N$ by $1/4$:
\[
\sum_{s=1}^\ell h_Y(Y_s) \le \frac{1}{4} h_N(Y) \qquad \text{and} \qquad \sum_{s=1}^\ell \rho_s \le \frac{1}{4} \rho.
\]
\end{enumerate}
Define a node's \emph{charge} as $\Phi(N,X,Y) := X + \alpha h_N(Y)$.
At a recursive node, where $m=Y$, the $\frac{3}{4} \alpha h_N(Y)$ saved from the shrinking outside-hint budget pays for all newly generated hints $O(h_N(Y))$ (Figure~\ref{fig:intro-hintsketch}):
\[
\sum_{s=1}^\ell \Phi(Y, X_s, Y_s) \le X + O\bigl(h_N(Y)\bigr) + \frac{1}{4} \alpha h_N(Y) \le \Phi(N,X,Y).
\]
After handling the boundary cases, we find that the geometric density contraction bounds the recursion depth by $O(\log\log n)$ and total nodes by $\polylog(n)$. The other costs are absorbed into the lower-order terms of Theorem~\ref{thm:hint-forest}.

\input{2609-version/fig-intro-hintsketch}

\subsubsection{Top-level sketch with \texorpdfstring{$dw+O(d)$}{dw + O(d)} length}

Recall that we use Pinsketch for row encoding only at the top level,
which we call the \emph{top-level sketch}.
In contrast, subsequent recursive layers ($\HintSketch$) use random linear sketches.
The intuition is to use Pinsketch for computational efficiency and random linear sketches for space efficiency.
In particular, Pinsketch supports polynomial-time encoding and decoding but uses more bits, while random linear sketches are more space-efficient, yet their expensive decoding restricts their use to the small subproblems created by the recursion.

At the top level, we use $m=d$ rows, each encoded into $B=O(\log\log n)$ Pinsketch words of $w$ bits.
By definition, the first $a-1$ words of a row's Pinsketch encoding already form a valid Pinsketch of capacity $a-1$ (Proposition~\ref{prop:pinsketch}).
Exploiting this property, worker $j$ updates its prediction at each word boundary $a \in [B]$ using the first $a-1$ recovered words (Figure~\ref{fig:intro-word-prediction}):
\begin{enumerate}
\item \textbf{Prefix decoding at word boundaries:} Before word $a$ begins, worker $j$ runs Pinsketch decoding on the first $a-1$ words to search for a candidate of size at most $a-1$. If no candidate exists, it outputs $\perp$ for all $w$ bits of word $a$.
\item \textbf{Predict until the first mismatch:} If a candidate $C$ is found, worker $j$ uses $C$ to predict the $w$ bits of word $a$ in order.
During the process, as soon as a predicted bit in word $a$ disagrees with the bit recovered by the manager, worker $j$ immediately discards $C$, outputs $\perp$ for the remaining bits of word $a$, and retries the prefix decoding at word $a+1$.
\item \textbf{When can predictions be wrong:} As long as $a > t_j$ (where $t_j = |S_j|$ is row $j$'s true weight), Pinsketch deterministically returns $C = S_j$, so worker $j$ predicts word $a$ and all subsequent words without error.
When $t_j \ge a$, however, the first $a-1$ words may occasionally match a false candidate $C \ne S_j$ of size $\le a-1$.
By bounding its probability, we can set parameters $(X,Y)$ and apply $\HintSketch$.
\end{enumerate}

\input{2609-version/fig-intro-word-prediction}

Next, we use the $\HintSketch$ guarantee discussed in Section~1.2.2 (Theorem~\ref{thm:hint-forest}) as a black box and show why the total space across all $\ell=Bw$ bit columns is
$dw+O(d)=L(n,d)+O(d)$.

\begin{itemize}
\item \textbf{Across words (hypergeometric row loads):}
For intuition, assume $\wt{x}=d$. Recall that $t_j$ is the number
of nonzeros in row $j$ and that only rows with $t_j\ge a$ can return $\perp$ or predict incorrectly in that word.
Thus, $q_a:=\sum_{j=1}^d\Pr[t_j\ge a]$ bounds the expected number of rows returning $\perp$ in each bit column of word $a$.
Since $\sum_{a\ge1}q_a=\sum_{j=1}^d\mathbb E[t_j]=d$, the total expected number of $\perp$ predictions across all columns is at most $w\sum_{a=1}^Bq_a\le dw=L(n,d)+O(d)$.

\item \textbf{Within a word (random scalar):}
We multiply each Pinsketch word by an independent random scalar $\alpha_{j,a} \in \F_{2^w}^*$ and have worker $j$ abort to $\perp$ for the rest of word $a$ after its first wrong bit.
Suppose a row with $t_j \ge a$ matches a false candidate $C \ne S_j$ and the predicted word is incorrect. This random scaling operation makes the $w$-bit discrepancy uniform on $\F_2^w \setminus \{0\}$.
Now, the first wrong bit $k \in [w]$ occurs with geometric probability $p_k \asymp 2^{-k}$, and aborting to $\perp$ prevents any second error in that word.
Thus, column $(a,k)$ has at most $q_a p_k$ expected wrong predictions (Figure~\ref{fig:intro-toplevel}, right) as only rows with $t_j\ge a$ can predict incorrectly.
Since both $(q_a/d)_a$ and $(p_k)_k$ have $O(1)$ Shannon entropy, the total expected error-location cost across all columns is
\[
\sum_{a=1}^B \sum_{k=1}^w h_d(q_a p_k) \le d\bigl(\log e + \mathsf{H}(q/d) + \mathsf{H}(p)\bigr) = O(d).
\]
\input{2609-version/fig-intro-toplevel}

\item \textbf{Concentration:}
While these bounds above control only expectations, we still need the numbers of rows returning $\perp$ or predicting incorrectly to concentrate.
We will show that an extension of McDiarmid's inequality (Lemma~\ref{lem:injection-concentration}) gives sub-Gaussian tails for both counts.
So adding small margins to the expectation bounds $q_a$ and $q_ap_k$ makes the column capacities sufficient with high probability, while adding only $o(d)$ bits in total.
\end{itemize}

Finally, some overloaded rows with $t_j>B$ may remain unrecovered.
We prove that, with high probability, these rows contain few nonzeros in total, so a final CR sketch recovers the residual using only $O(d)$ additional bits.

\subsection{Related Work}
\label{sec:intro-related}

\paragraph{Algebraic vs.\ peeling sketches.}
Algebraic sketches via characteristic polynomials~\cite{MTZ03} and BCH-based Pinsketch~\cite{DORS08} use about $d$ field elements and $O(d)$ field operations per update.
PBS~\cite{PBS20} combines hash partitioning with BCH coding.
Hashing-based IBLTs~\cite{GM11} and their refinements via XOR-only cells~\cite{HPW23}, algebraic repair~\cite{BKW24}, limited independence~\cite{FLOS24}, and rateless/zero-error variants~\cite{YGA24,MBYR23,KYR25} achieve fast updates and peeling decoding, but still require $\ge d\log n$ bits.
Recently, combining high-capacity cells with spatial coupling~\cite{KRU11,Wal25}, XYZ-Sketch~\cite{XYZ26} approaches $d\log n$ bits, while Stuffed IBLTs~\cite{KPW26} achieve $(1+\delta)L(n,d)+O(d)$ bits.
Their analysis~\cite[Theorem~1 and Sections~4.4--4.6]{KPW26} requires $\delta$ to be a fixed constant, so it does not give $L(n,d)+O(d)$ bits. We eliminate the multiplicative slack $\Theta(\delta d\log(n/d))$ to reach $L(n,d) + O(d)$ bits, while using $\polylog(n)$ updates and $d n^{o(1)}$ decoding.

\paragraph{Relation to error-correcting codes.}
Linear set reconciliation is closely related to syndrome decoding.
For sets $A,B\subseteq[n]$, linearity gives
$H\mathbf{1}_A+H\mathbf{1}_B=H\mathbf{1}_{A\triangle B}$.
Similarly, for a codeword $c\in\ker H$ corrupted by an error vector $e$, the received word $y=c+e$ satisfies $Hy=He$, so recovering $e$ from this syndrome recovers $c$ as well.
Our sketch therefore supplies a syndrome decoder for $\ker H$.
However, a classical code must correct every error pattern $e$ of weight at most $d$, whereas our guarantee fixes $e$ before the sketch is sampled and allows failure probability $\eps$ over that randomness.
In particular, our result does not give a worst-case $d$-error-correcting code with redundancy $L(n,d)+O(d)$.

\paragraph{Polar coding and successive recovery.}
Polar coding has also been applied to sparse recovery~\cite{Abb10}, but in a different setting: the input coordinates are independent and identically distributed, whereas our guarantee holds for every fixed sparse input $x$ with $\wt{x}\le d$.
For binary i.i.d.\ sources with fixed density $d/n\in(0,1/2)$, Abbe's polar construction~\cite[Theorem~4 and its proof]{Abb10} achieves length $L(n,d)+o(n)$, error probability $O(2^{-n^\beta})$ for any fixed $0<\beta<1/2$, and $O(n\log n)$ encoding and decoding time.
These bounds are asymptotic in $n$ for a fixed density, so they do not give $O(d)$ redundancy when $d/n\to0$; moreover, encoding and decoding take $O(n\log n)$ time, compared with our $\log^{O(1)}n$ update time and $dn^{o(1)}$ decoding time.
Whether our approach can be adapted to their source-coding setting remains an interesting question for future work.

\paragraph{Asymmetric information and sparse recovery.}
Problem~\ref{prob:intro-hinted} specializes asymmetric document exchange~\cite{CL20}. In our construction, the hint $J_s$ consists of the indices of rows whose decoders return $\perp$ when predicting column $s$. For residual cleanup, we apply the deterministic expander sketch of Cheraghchi and Ribeiro~\cite{CR23}, whose $O(t\log^2 N)$ bit cost is negligible on tiny residuals.

\subsection{Paper Organization}
\label{sec:paper-organization}
Section~\ref{sec:preliminaries} introduces notation and the sketches used as building blocks.
Section~\ref{sec:construction} describes our construction, and
Section~\ref{sec:analysis-framework} develops the probabilistic tools and decoding framework.
Section~\ref{sec:analysis-hint} analyzes $\HintSketch$.
Section~\ref{sec:analysis-top} applies this result to the top-level sketch to prove the main theorem.
Section~\ref{sec:ai-methodology} describes our use of AI tools.
The proofs of the probabilistic tools appear in Appendix~\ref{app:probabilistic-tools}.

%% file: 2609-version/fig-intro-architecture.tex
\begingroup
\colorlet{introText}{black}
\definecolor{introLine}{RGB}{83,101,117}
\definecolor{introFill}{RGB}{230,237,243}
\definecolor{introHint}{RGB}{237,233,222}
\definecolor{introError}{RGB}{239,215,203}
\definecolor{introErrorLine}{RGB}{140,75,46}
\definecolor{introColumn}{RGB}{228,237,229}
\definecolor{introColumnLine}{RGB}{86,120,110}
\definecolor{introIntersection}{RGB}{211,226,221}
\begin{figure}[!htbp]
\edrev
\centering
\begin{tikzpicture}[x=1cm,y=1cm,font=\figurefontsize,text=introText,
  box/.style={draw=introLine,rounded corners=2pt,line width=.65pt,
    fill=white,align=center,inner sep=5pt},
  arr/.style={-{Latex[length=2mm]},draw=introLine,line width=.8pt}]
\path[use as bounding box] (0,-1.15) rectangle (15.9,4.9);
\node[anchor=west,font=\figurefontsize\bfseries] at (0,4.65) {(a) Encoding};

\node (input) at (.35,1.625) {$x$};
\foreach \y/\j in {2.925/1,1.625/j,.325/m}{
  \node (row\j) at (2,\y) {$x_{\j}$};
  \draw[arr] (input.east)--(row\j.west);
  \draw[arr] (2.3,\y)--(4.85,\y)
    node[midway,above=2pt] {$\mathsf{Enc}_{\j}$};
}
\foreach \y in {2.275,.975}{
  \node at (2,\y) {$\vdots$};
}

\fill[introFill] (5,1.3) rectangle (10.4,1.95);
\fill[introColumn] (7.7,0) rectangle (8.6,3.25);
\fill[introIntersection] (7.7,1.3) rectangle (8.6,1.95);
\draw[introLine,line width=.65pt] (5,0) rectangle (10.4,3.25);
\foreach \x in {5.9,6.8,7.7,8.6,9.5}
  \draw[introLine,line width=.4pt] (\x,0)--(\x,3.25);
\foreach \y in {.65,1.3,1.95,2.6}
  \draw[introLine,line width=.4pt] (5,\y)--(10.4,\y);
\draw[introColumnLine,line width=1pt,rounded corners=3pt]
  (7.7,0) rectangle (8.6,3.25);
\node[fill=introFill,inner sep=3pt] at (6.35,1.625)
  {$\mathsf{Enc}_j(x_j)$};

\node[box,minimum width=3.35cm,minimum height=1.65cm] (sketch) at (13.7,1.625)
  {$\bigl(H_sC_s(x)\bigr)_{s=1}^{\ell}$\\[7pt]$\CR_{n,t}x$};
\node at (13.7,2.95) {Stored sketch};
\foreach \x in {5.45,6.35,7.25,8.15,9.05,9.95}
  \draw[introLine,line width=.55pt] (\x,3.25)--(\x,4.05);
\foreach \x/\colLabel in {5.45/{C_1(x)},6.8/{\cdots},9.05/{\cdots},9.95/{C_\ell(x)}}
  \node[fill=white,inner sep=2pt] at (\x,3.65) {$\colLabel$};
\node[fill=white,inner sep=2pt] at (8.15,3.65) {$C_s(x)$};
\draw[arr] (5.45,4.05)--(13.7,4.05)--(sketch.north);
\node[fill=white,inner sep=2pt] at (11.8,4.05) {Sketch each column};
\draw[arr] (input.south)--(.35,-.75)--(13.7,-.75)--(sketch.south);
\node[fill=white,inner sep=3pt] at (7.2,-.75)
  {$\CR_{n,t}x$ for residual recovery};
\end{tikzpicture}

\medskip
\begin{tikzpicture}[x=1cm,y=1cm,font=\figurefontsize,text=introText,
  box/.style={draw=introLine,rounded corners=2pt,line width=.65pt,
    fill=white,align=center,inner sep=5pt},
  arr/.style={-{Latex[length=1.8mm]},draw=introLine,line width=.65pt}]
\path[use as bounding box] (0,-4.65) rectangle (15.9,2.0);
\node[anchor=west,font=\figurefontsize\bfseries] at (0,1.75) {(b) Decoding column $s$};
\node at (2.2,1.2) {$m\times\ell$ matrix};
\node at (5.3,1.2) {Predictions};
\node at (10.2,1.2) {Stored sketch};
\node at (13.6,1.2) {Recovered column};
\node at (5.3,.5) {$q_{j,s}$};
\node at (7.6,.5) {$P_s$};
\node at (13.6,.5) {$C_s(x)$};

\fill[introFill] (.4,-2.35) rectangle (2.2,.05);
\fill[introHint] (2.2,-2.35) rectangle (2.8,.05);
\draw[introLine,line width=.65pt] (.4,-2.35) rectangle (4,.05);
\foreach \x in {1,1.6,2.2,2.8,3.4}
  \draw[introLine,line width=.4pt] (\x,-2.35)--(\x,.05);
\foreach \y in {-.55,-1.15,-1.75}
  \draw[introLine,line width=.4pt] (.4,\y)--(4,\y);
\draw[introLine,line width=1.1pt] (2.2,-2.35) rectangle (2.8,.05);
\draw[introLine,decorate,decoration={brace,amplitude=3pt}]
  (.4,.22)--(2.2,.22) node[midway,above=4pt,text=introText] {$1,\ldots,s-1$};
\node at (2.5,.5) {$s$};
\node at (3.4,.5) {unread};
\foreach \y/\j/\a/\b/\c in {-.25/1/0/1/0,-.85/2/1/0/1,-1.45/3/0/0/1,-2.05/4/1/1/0}{
  \node[anchor=east,font=\figurefontsize] at (.3,\y) {$\j$};
  \node at (.7,\y) {$\a$};
  \node at (1.3,\y) {$\b$};
  \node at (1.9,\y) {$\c$};
  \node at (2.5,\y) {$?$};
  \draw[arr] (4.15,\y)--(4.8,\y);
}
\node[font=\figurefontsize] at (1.3,-2.65) {recovered};
\fill[introHint] (4.95,-1.15) rectangle (5.65,-.55);
\fill[introHint] (4.95,-2.35) rectangle (5.65,-1.75);
\draw[introLine,line width=.65pt] (4.95,-2.35) rectangle (5.65,.05);
\foreach \y in {-.55,-1.15,-1.75}
  \draw[introLine,line width=.4pt] (4.95,\y)--(5.65,\y);
\foreach \y/\q in {-.25/1,-.85/\perp,-1.45/0,-2.05/\perp}
  \node at (5.3,\y) {$\q$};

\fill[introHint] (7.25,-1.15) rectangle (7.95,-.55);
\fill[introHint] (7.25,-2.35) rectangle (7.95,-1.75);
\draw[introLine,line width=.65pt] (7.25,-2.35) rectangle (7.95,.05);
\foreach \y in {-.55,-1.15,-1.75}
  \draw[introLine,line width=.4pt] (7.25,\y)--(7.95,\y);
\foreach \y/\p in {-.25/1,-.85/0,-1.45/0,-2.05/0}
  \node at (7.6,\y) {$\p$};
\draw[arr] (5.65,-1.15)--(7.1,-1.15)
  node[midway,above=5pt] {$\perp\to0$};

\node[box,fill=introFill,minimum width=2.6cm,minimum height=1.1cm] (decode)
  at (10.2,-1.15) {\HintSketch\\[3pt]recover $D_s$};
\node (stored) at (10.2,.5) {$H_sC_s(x)$};
\draw[arr] (stored.south)--(decode.north);
\draw[arr] (7.95,-1.15)--(decode.west);

\node[box,fill=introHint,minimum width=3cm] (hint) at (5.3,-3.3)
  {Hint $J_s=\{2,4\}$};
\draw[arr] (5.3,-2.35)--(hint.north);
\draw[arr] (hint.east)--(10.2,-3.3)--(decode.south);

\draw[arr] (decode.east)--(13.1,-1.15)
  node[midway,above=5pt] {$D_s+P_s$};
\fill[introHint] (13.25,-1.15) rectangle (13.95,-.55);
\fill[introError] (13.25,-1.75) rectangle (13.95,-1.15);
\fill[introHint] (13.25,-2.35) rectangle (13.95,-1.75);
\draw[introLine,line width=.85pt] (13.25,-2.35) rectangle (13.95,.05);
\foreach \y in {-.55,-1.15,-1.75}
  \draw[introLine,line width=.4pt] (13.25,\y)--(13.95,\y);
\draw[introErrorLine,line width=.9pt] (13.25,-1.75) rectangle (13.95,-1.15);
\foreach \y/\v in {-.25/1,-.85/0,-1.45/1,-2.05/1}
  \node at (13.6,\y) {$\v$};
\foreach \y/\annotation in {-.85/{\perp\to0},-1.45/{0\to1},-2.05/{\perp\to1}}{
  \draw[introLine,line width=.4pt] (14.1,\y)--(14.45,\y);
  \node[anchor=west] at (14.55,\y) {$\annotation$};
}
\draw[arr] (13.6,-2.35)--(13.6,-4.35)--(2.5,-4.35)--(2.5,-2.35);
\node[fill=white,inner sep=3pt] at (8.05,-4.35)
  {Fill in column $s$, then continue with column $s+1$.};
\end{tikzpicture}
\caption{Row encoding and column recovery.
(a) Independent row encodings form the matrix.
Its column sketches and a separate CR sketch of $x$ form the stored output.
(b) Columns $1,\ldots,s-1$ have been recovered, and column $s$ is being decoded.
Here $J_s=\{2,4\}$ marks the missing predictions. Replacing $\perp$ by $0$ gives $P_s=(1,0,0,0)^{\top}$.
\HintSketch\ then recovers $D_s=C_s(x)+P_s$ from $H_sC_s(x)+H_sP_s$ and $J_s$.
Adding back $P_s$ fills the missing entries and corrects row $3$'s prediction.}
\label{fig:intro-architecture}
\end{figure}
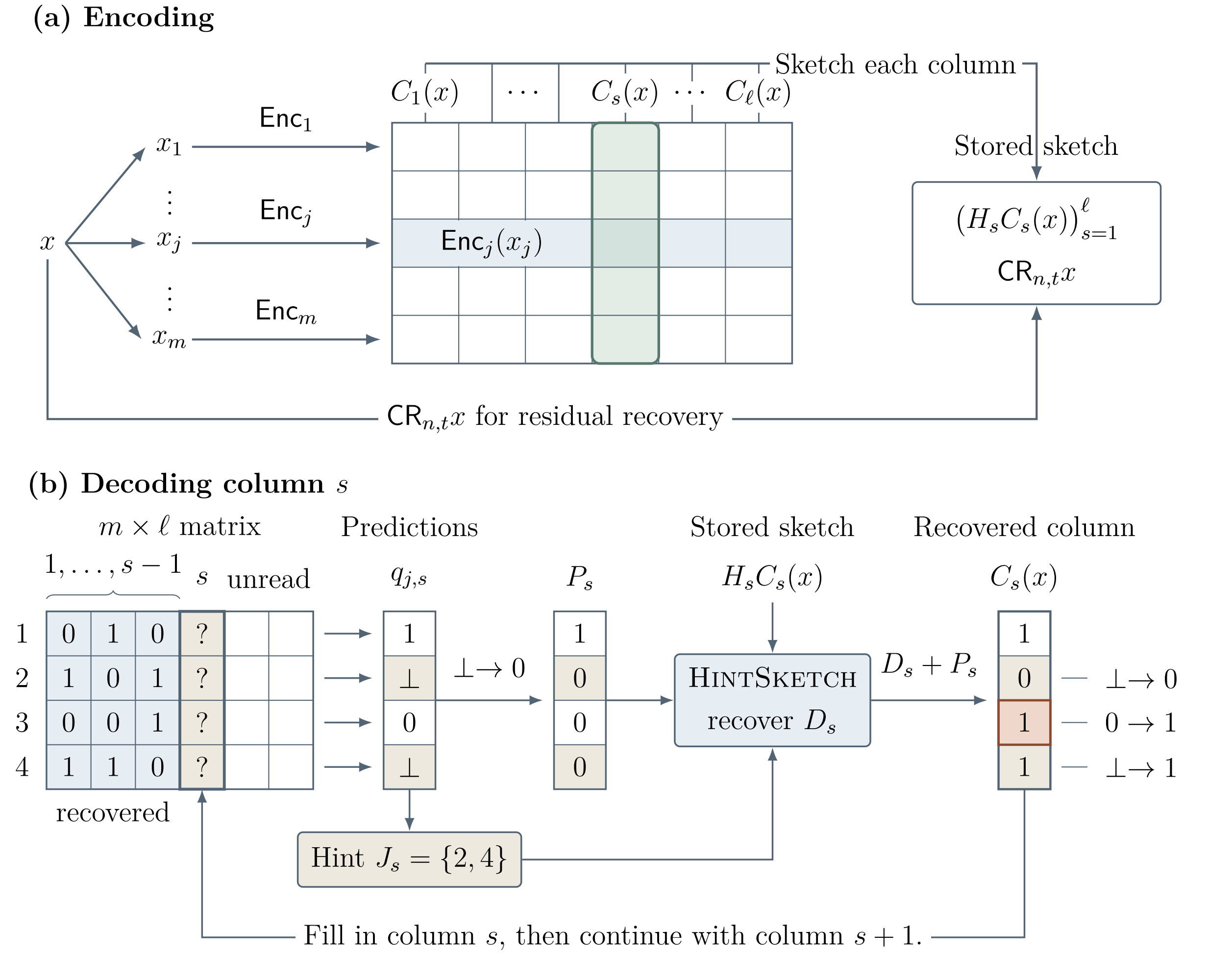
\endgroup

%% file: 2609-version/fig-intro-hintsketch.tex
\begingroup
\colorlet{introText}{.}
\definecolor{introLine}{RGB}{83,101,117}
\definecolor{introFill}{RGB}{238,238,238}
\definecolor{introHint}{RGB}{237,233,222}
\definecolor{introOutside}{RGB}{188,208,226}
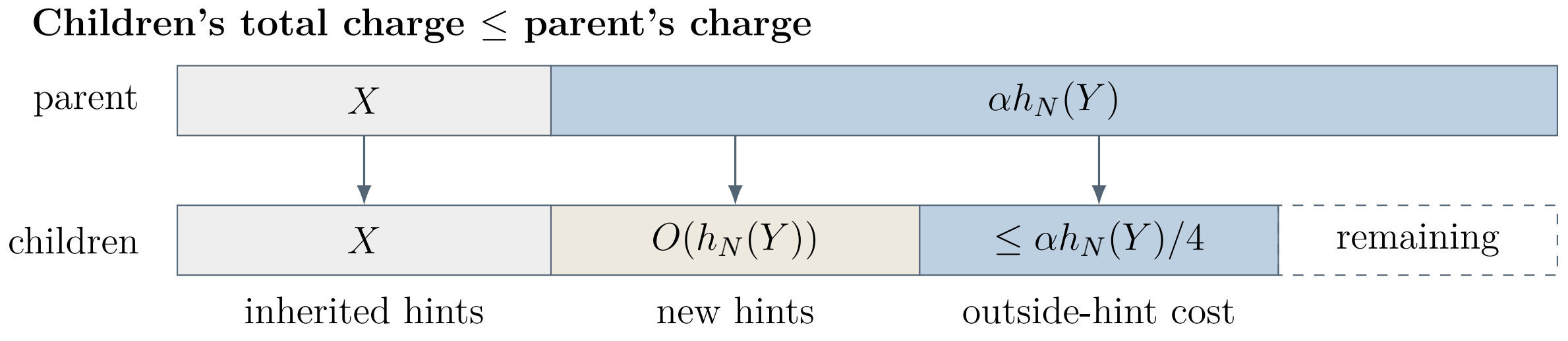
\begin{figure}[!htbp]
\centering
\begin{tikzpicture}[x=1cm,y=1cm,font=\figurefontsize,text=introText,
  arr/.style={-{Latex[length=2mm]},draw=introLine,line width=.6pt}]
\node[anchor=west,font=\figurefontsize\bfseries] at (0,.35) {Children's total charge $\le$ parent's charge};
\node[anchor=east] at (1.35,-.4) {parent};
\filldraw[draw=introLine,fill=introFill] (1.6,-.75) rectangle (5.35,-.05);
\filldraw[draw=introLine,fill=introOutside] (5.35,-.75) rectangle (15.45,-.05);
\node at (3.475,-.4) {$X$};
\node at (10.4,-.4) {$\alpha h_N(Y)$};
\node[anchor=east] at (1.35,-1.8) {children};
\filldraw[draw=introLine,fill=introFill] (1.6,-2.15) rectangle (5.35,-1.45);
\filldraw[draw=introLine,fill=introHint] (5.35,-2.15) rectangle (9.05,-1.45);
\filldraw[draw=introLine,fill=introOutside] (9.05,-2.15) rectangle (12.65,-1.45);
\draw[draw=introLine,dashed] (12.65,-2.15) rectangle (15.45,-1.45);
\node at (3.475,-1.8) {$X$};
\node at (7.2,-1.8) {$O(h_N(Y))$};
\node at (10.85,-1.8) {$\le \alpha h_N(Y)/4$};
\node at (14.05,-1.8) {remaining};
\node at (3.475,-2.5) {inherited hints};
\node at (7.2,-2.5) {new hints};
\node at (10.85,-2.5) {outside-hint cost};
\draw[arr] (3.475,-.75)--(3.475,-1.45);
\draw[arr] (7.2,-.75)--(7.2,-1.45);
\draw[arr] (10.85,-.75)--(10.85,-1.45);
\end{tikzpicture}
\caption{The decrease in outside-hint cost pays for the new hints, so the total charge $X+\alpha h_N(Y)$ does not increase.}
\label{fig:intro-hintsketch}
\end{figure}
\endgroup

%% file: 2609-version/fig-intro-word-prediction.tex
\begingroup
\definecolor{wordLine}{RGB}{99,108,116}
\definecolor{wordCorrect}{RGB}{230,239,226}
\definecolor{wordError}{RGB}{239,215,203}
\definecolor{wordErrorLine}{RGB}{169,101,76}
\definecolor{wordMissing}{RGB}{240,240,238}
\definecolor{wordRecovered}{RGB}{231,237,242}
\begin{figure}[!htbp]
\edrev
\colorlet{wordText}{.}
\centering
\begin{tikzpicture}[x=.97cm,y=.65cm,font=\figurefontsize,text=wordText,
  arrow/.style={-{Latex[length=1.7mm]},draw=wordLine,line width=.65pt},
  cell/.style={draw=wordLine,line width=.45pt}]
\path[use as bounding box] (0,-.15) rectangle (16,7.95);
\node[anchor=west,font=\figurefontsize\bfseries] at (0,7.65) {At the word boundary};
\node[font=\figurefontsize\bfseries] at (8.4,7.65) {Predictions in word $a$};
\node[font=\figurefontsize\bfseries] at (14.05,7.65) {Next word};
\foreach \i/\k in {0/1,1/2,2/3,3/{\cdots},4/{w-1},5/w}
  \node at ({5.9+\i},6.95) {$\k$};
\foreach \y in {6.65,5.35,1.6,-.1}
  \draw[wordLine!50,line width=.4pt] (0,\y)--(16,\y);

\node[anchor=west,font=\figurefontsize\bfseries] at (0,6.05) {(i) No candidate};
\foreach \i in {0,...,5}{
  \filldraw[cell,fill=wordMissing]
    ({5.4+\i},5.55) rectangle ({6.4+\i},6.4);
  \node at ({5.9+\i},5.975) {$\perp$};
}
\draw[arrow] (11.6,5.975)--(12.05,5.975);
\node[align=center] at (14.05,5.975)
  {Retry\\prefix decoding};

\begin{scope}[yshift=-3mm]
\node[anchor=west,font=\figurefontsize\bfseries] at (0,4.6) {(ii) First mismatch};
\node[anchor=west] at (0,3.8) {Candidate $C$};
\node[anchor=east] at (5.15,3.95) {Predicted};
\node[anchor=east] at (5.15,2.85) {Recovered};
\foreach \i/\v/\shade in {
  0/1/wordCorrect,1/0/wordCorrect,2/1/wordError,
  3/\perp/wordMissing,4/\perp/wordMissing,5/\perp/wordMissing}{
  \filldraw[cell,fill=\shade]
    ({5.4+\i},3.525) rectangle ({6.4+\i},4.375);
  \node at ({5.9+\i},3.95) {$\v$};
}
\draw[wordLine,decorate,decoration={brace,amplitude=4pt}]
  (8.4,4.6)--(11.4,4.6) node[midway,above=7pt,text=wordText] {Return $\perp$};
\foreach \i/\v in {0/1,1/0,2/0,3/1,4/0,5/1}{
  \filldraw[cell,fill=wordRecovered]
    ({5.4+\i},2.425) rectangle ({6.4+\i},3.275);
  \node at ({5.9+\i},2.85) {$\v$};
}
\draw[wordErrorLine,line width=.9pt] (7.4,2.425) rectangle (8.4,3.275);
\draw[wordErrorLine,line width=.65pt] (7.9,3.275)--(7.9,3.525);
\draw[arrow] (11.6,3.95)--(12.05,3.95);
\node[align=center] at (14.05,3.95)
  {Retry\\prefix decoding\\at word $a+1$};
\end{scope}

\node[anchor=west,align=left,font=\figurefontsize\bfseries] at (0,.7)
  {(iii) Support recovered};
\foreach \i/\v in {0/1,1/0,2/0,3/1,4/0,5/1}{
  \filldraw[cell,fill=wordCorrect]
    ({5.4+\i},.275) rectangle ({6.4+\i},1.125);
  \node at ({5.9+\i},.7) {$\v$};
}
\draw[arrow] (11.6,.7)--(12.05,.7);
\node[align=center] at (14.05,.7)
  {Correct predictions\\in all later words};
\end{tikzpicture}
\caption{Three illustrative cases of word-by-word prediction for one row.
Prefix decoding uses the first $a-1$ recovered words.
In (ii), the first mismatch ends prediction for the current word.}
\label{fig:intro-word-prediction}
\end{figure}
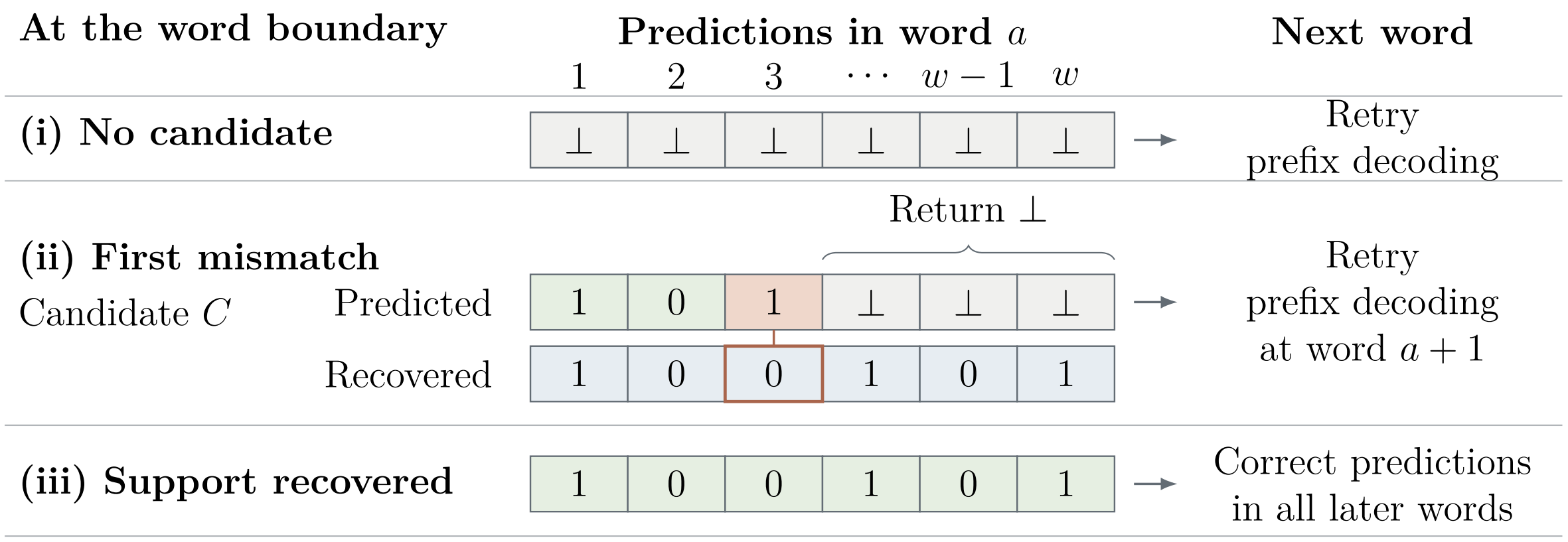
\endgroup

%% file: 2609-version/fig-intro-toplevel.tex
\begingroup
\colorlet{introText}{.}
\definecolor{introLine}{RGB}{142,151,160}
\definecolor{introHint}{RGB}{68,119,170}
\definecolor{introAccent}{RGB}{45,85,125}
\begin{figure}[!htbp]
\centering
\begin{tikzpicture}[x=1cm,y=1cm,font=\figurefontsize,text=introText,>=Stealth]
\node[font=\figurefontsize\bfseries] at (2.65,3.3) {Across words};
\foreach \lab/\x/\h in {1/0/1.85,2/1.8/1.0,a/4.0/.52}{
  \filldraw[fill=introHint!5,draw=introLine,line width=.5pt,rounded corners=2pt]
    (\x,.2) rectangle ({\x+1.5},{.4+\h});
  \foreach \k in {1,...,6}{
    \filldraw[fill=introHint!15,draw=introLine!70,line width=.25pt]
      ({\x+.1+.22*(\k-1)},.3) rectangle ({\x+.28+.22*(\k-1)},{.3+\h});
    \fill[introHint] ({\x+.1+.22*(\k-1)},.3)
      rectangle ({\x+.28+.22*(\k-1)},{.3+(32/63)*\h*pow(.5,\k-1)});
  }
  \node at ({\x+.75},{.7+\h}) {$q_{\lab}$};
  \node at ({\x+.75},-.15) {word $\lab$};
}
\node at (3.65,.7) {$\cdots$};
\draw[introAccent,line width=.8pt,rounded corners=2pt] (4,.2) rectangle (5.5,.92);
\draw[introLine,densely dashed] (5.5,.92)--(6.55,3.72);
\draw[introLine,densely dashed] (5.5,.2)--(6.55,-.55);
\node[align=center] at (2.75,-.85)
  {$q_a=\sum_j\Pr[T_j\ge a]$\\hypergeometric row loads};
\filldraw[fill=white,draw=introAccent,line width=.8pt,rounded corners=2pt]
  (6.55,-.55) rectangle (15.45,3.72);
\node[font=\figurefontsize\bfseries] at (11,3.3) {Inside word $a$};
\foreach \i/\lab in {0/1,1/2,2/3,3/{\cdots},4/k,5/w}{
  \filldraw[fill=introHint!15,draw=introLine,line width=.4pt]
    ({7.05+1.32*\i},.05) rectangle ({7.81+1.32*\i},2.15);
  \fill[introHint] ({7.05+1.32*\i},.05)
    rectangle ({7.81+1.32*\i},{.05+2.1*(32/63)*pow(.5,\i)});
  \node at ({7.43+1.32*\i},-.23) {$\lab$};
}
\node (hintlabel) at (11,2.65) {no prediction: $\mathbb E[\perp]\le q_a$};
\draw[->,introAccent] (hintlabel.south)--(10.1,2.15);
\node[fill=white,inner sep=4pt,align=center] (errlabel) at (12.55,1.25)
  {incorrect prediction\\$\mathbb E[\mathrm{incorrect}]\le q_ap_k$};
\draw[->,introAccent] (errlabel.west)--(10.1,{.05+2.1*8/63});
\node at (11,-.85) {$p_k\asymp2^{-k}$};
\end{tikzpicture}
\caption{Two scales of the top-level space bound. Light columns show the bound $q_a$ on the expected number of rows returning $\perp$, while dark bars show the bound $q_ap_k$ on incorrect predictions. Across words, $q_a$ follows the hypergeometric load tail. Within a word, random field scaling gives geometric decay; after its first wrong prediction, a worker returns $\perp$ for the rest of that word.}
\label{fig:intro-toplevel}
\end{figure}
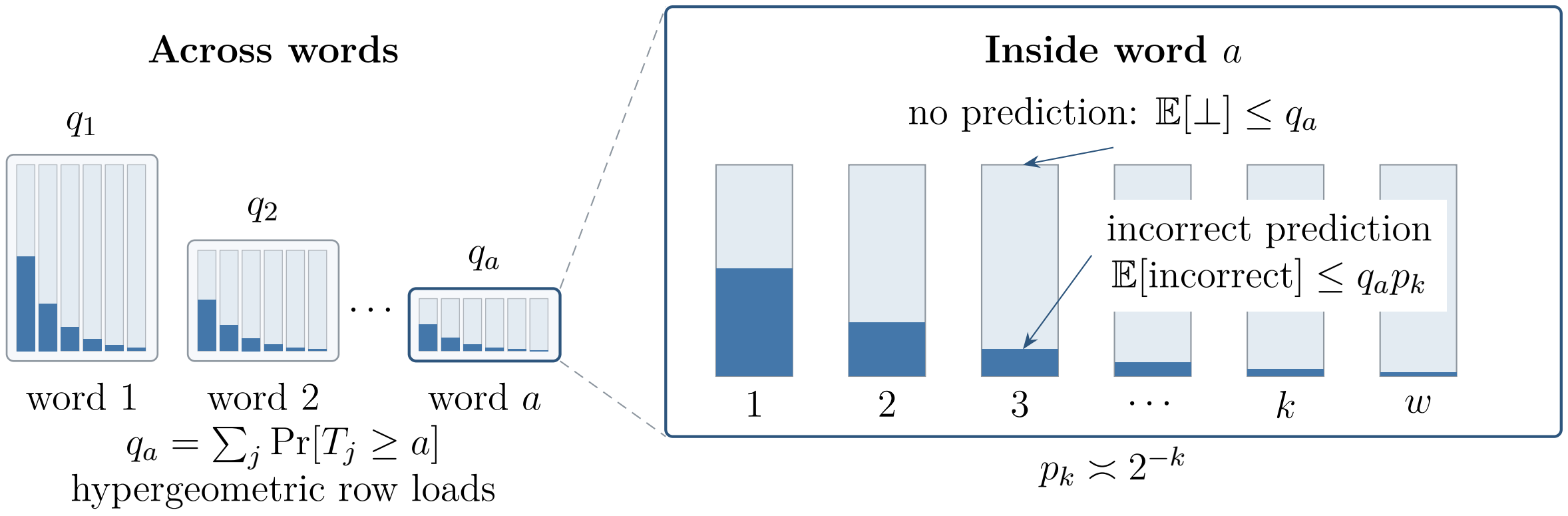
\endgroup

%% file: 2609-version/2preliminaries.tex
\section{Preliminaries}
\label{sec:preliminaries}

\subsection{Notations and conventions}
\label{sec:notations}

Throughout the paper, all logarithms are to base two unless written as $\ln$. For an integer
$N\ge1$, write $[N]=\{1,\ldots,N\}$.
For sets $A,B$, write $A\triangle B=(A\setminus B)\cup(B\setminus A)$
for their symmetric difference.
We identify subsets $A\subseteq[N]$ with
their characteristic vectors $\mathbf{1}_A\in\F_2^N$, defined by
\[
(\mathbf{1}_A)_i=
\begin{cases}
1, & i\in A,\\
0, & i\in[N]\setminus A.
\end{cases}
\]
Thus addition corresponds to
symmetric difference. For $x\in\F_2^N$, write $\supp(x)=\{i:x_i=1\}$,
$\wt{x}=|\supp(x)|$, and let $e_i$ be the $i$th standard basis vector.

For integers $0\le t\le N$, define
\begin{equation}
\mathcal S_{N,t}:=\{x\in\F_2^N:\wt{x}\le t\},
\qquad L(N,t):=\log|\mathcal S_{N,t}|
 =\log\sum_{a=0}^{t}\binom Na.
\label{eq:counting-benchmark}
\end{equation}
For convenience, we assume $n\ge2$, $2\le d\le n$, and $0<\eps\le1/2$; we write
$\lambda:=\ceil{\log(1/\eps)}$.
We use $n$ for the original universe size and $N$ for a local universe.
Unless stated otherwise, constants in asymptotic notation are absolute.

A public-coin binary linear sketch is a family of maps
$H_r:\F_2^n\to\F_2^m$, indexed by public randomness $r$.
We call its maximum column weight $\max_i\wt{H_re_i}$ its column locality, and toggling coordinate $i$ adds $H_re_i$ to the sketch. All probability guarantees are for each fixed input chosen independently of $r$. Running times are measured on a word-RAM with $\Theta(\log n)$-bit words.
We assume $O(1)$-time access to public random permutations and their inverses, random coefficients and precomputed expander adjacency tables used in the CR sketch.
Their storage is excluded from the sketch length, and preprocessing is excluded from the reported running times.

We write $\mathsf H$ for Shannon entropy. Specifically, for a discrete
distribution $p=(p_i)_i$, $\mathsf H(p)=\sum_i p_i\log(1/p_i)$,
with zero terms interpreted as zero. For a discrete random variable
$T$ with distribution $p$, we write $\mathsf H(T)=\mathsf H(p)$.
For $z\in (0,N]$, define $h_N(z)=z\log(eN/z)$, with $h_N(0)=0$.
This approximates the information needed to specify a support of size
at most $z$ in $[N]$: for integer $1\le t\le N/2$,
$L(N,t)\le h_N(t)=L(N,t)+O(t)$.

\subsection{Tools from prior work}
\label{sec:prior-tools}

\paragraph{Pinsketch.}
We use the BCH-based \textsc{Pinsketch} of Dodis, Ostrovsky, Reyzin,
and Smith~\cite[Section~6.3]{DORS08}. Fix
$w=\ceil{\log(N+1)}$, a basis of $\F_{2^w}$ over $\F_2$, and an injection
$\iota:[N]\hookrightarrow\F_{2^w}^{*}$ with an efficiently computable inverse
on its image.\footnote{For example, map each $i\in[N]$ to the field element whose coordinates in the chosen basis are the $w$-bit binary representation of $i$.} For $a\ge1$, put
$P_{a}(x):=\sum_{i\in\supp(x)}\iota(i)^{a}$.

\begin{proposition}[Pinsketch, \cite{DORS08}]
\label{prop:pinsketch}
For every $0\le t\le N$, the map
\begin{equation}
\operatorname{Pin}_{N,t}(x):=(P_{2j-1}(x))_{j=1}^{t}=(P_1(x),P_3(x),\cdots,P_{2t-1}(x))
\label{eq:pinsketch}
\end{equation}
is deterministic and $\F_2$-linear, uses exactly $tw$ bits, and permits
deterministic recovery of every $x\in\mathcal S_{N,t}$ in time
$\poly(t,w)$. A coordinate update uses $O(t)$ field operations and
changes at most $tw$ bits.

On any input syndrome, the decoder runs in $\poly(t,w)$ time and returns an $x\in\mathcal S_{N,t}$ matching that syndrome, or $\perp$ if no such $x$ exists.
Moreover, for every $0\le a\le t$, the first $a$ field elements of
$\operatorname{Pin}_{N,t}(x)$ form $\operatorname{Pin}_{N,a}(x)$.
\end{proposition}

\begin{remark}
Using the Berlekamp trace algorithm for root finding gives a
deterministic Pinsketch decoder running in $\poly(t,w)$ time~\cite[Section~4.2, Algorithm~1]{Str12}.
\end{remark}

\paragraph{The CR sketch.}
We call the binary linear sketch of Cheraghchi and Ribeiro the
\emph{CR sketch}. Their construction~\cite[Theorems~5 and~11]{CR23}
gives the following result.

\begin{proposition}[Deterministic sparse recovery, \cite{CR23}]
\label{prop:cr-sketch}
For every $N\ge1$ and $0\le t\le N$, there is a fixed binary matrix
$\mathsf{CR}_{N,t}$ with $O(t\log^2(N+2))$ rows and maximum column weight $O(\log^2(N+2))$.
A coordinate update of $x$ takes $O(\log^2(N+2))$ time. 
Given $\mathsf{CR}_{N,t}x$, a deterministic decoder recovers every
$x\in\mathcal S_{N,t}$ in time $O(t\log(t+2)\log(N+2))$.
\end{proposition}

Their construction combines a bipartite expander with bitmasking,
which records the binary indices of neighboring coordinates to enable
efficient recovery. We fix a suitable expander during preprocessing, hence the resulting sketch supports deterministic decoding.

%% file: 2609-version/3construction.tex
%

\section{Construction}
\label{sec:construction}

\subsection{General framework}
\label{sec:construction-framework}

\paragraph{Hinted sparse recovery problem.}
Our construction uses a variant of sparse recovery in which the decoder
receives additional information (named \emph{hint}) about the support.
We now formalize this problem.

\begin{problem}[Hinted sparse recovery]
\label{prob:hinted-recovery}
Let $N\ge1$ and $0\le X,Y\le N$ be integers. The decoder is given
a syndrome $Hx$ of a vector $x\in\F_2^N$ and a \textbf{hint} $J\subseteq[N]$
satisfying
\begin{equation}
|J|\le X,\qquad |\supp(x)\setminus J|\le Y.
\label{eq:hint-promise}
\end{equation}
The goal is to recover $x$ from $Hx$ and $J$.
\end{problem}

Intuitively, $J$ contains all but at most $Y$ nonzero coordinates of $x$,
without specifying their values.
The hint is available only to the decoder; the sketching matrix is
chosen independently of $J$.

\begin{remark}
Hinted sparse recovery is a special case of asymmetric document
exchange~\cite{CL20}: the decoder knows a set containing all but at most
$Y$ elements of the input support. Our construction requires control of
the leading constants in the sketch length beyond that provided by a
direct application of the general result.
\end{remark}

This problem arises when recovering a vector $y$ from partial predictions.
Specifically, we predict some coordinates of a vector and leave the others unspecified. Let $J$ be the set of unspecified coordinates and $\hat{y}$ the prediction vector, with zeros on $J$.
Recovering the vector therefore reduces to hinted recovery of $y-\hat{y}$, whose nonzero coordinates outside $J$ correspond exactly to incorrect predictions.
Intuitively, on $J$ we \emph{ know that we do not know the values}; outside $J$,
we \emph{have predictions but do not know which are incorrect}.

This problem will be solved in Section~\ref{sec:hintsketch-structure}, where we construct a sketch called \HintSketch.
For the purpose of describing the general framework, we may temporarily treat it as a black box.

\begin{figure}[htbp]
\centering
\begin{tikzpicture}[
  font=\figurefontsize,
  arr/.style={
    -{Latex[length=2.5mm,width=1.8mm]},
    thick
  },
  manager/.style={
    draw,rounded corners=2pt,align=center,inner sep=6pt
  }
]
\draw (0,0) rectangle (2.2,2.4);
\foreach \y in {0.6,1.2,1.8}
  \draw (0,\y)--(2.2,\y);
\foreach \y/\v in {2.1/x_1,1.5/x_2,1/\vdots,0.3/x_m}
  \node at (1.1,\y) {$\v$};
\node at (1.1,3.35) {Input: $m\times N$};
\node[font=\figurefontsize] at (1.1,-0.35)
  {$x_j\in\mathbb F_2^N$};

\foreach \y/\j in {2.1/1,1.5/2,0.3/m}
  \draw[arr] (2.3,\y)--(4.2,\y)
    node[midway,above,font=\figurefontsize] {$\mathsf{Enc}_{\j}$};
\node at (3.25,1.1) {$\vdots$};

\begin{scope}[shift={(4.3,0)}]
  \draw (0,0) rectangle (3.6,2.4);
  \foreach \x in {0.6,1.2,1.8,2.4,3.0}
    \draw (\x,0)--(\x,2.4);
  \foreach \y in {0.6,1.2,1.8}
    \draw (0,\y)--(3.6,\y);

  \foreach \x/\a/\b/\c in {
    0.3/0/1/1,0.9/1/0/1,1.5/0/1/0,
    2.1/1/0/0,2.7/0/1/1,3.3/1/1/0} {
    \node at (\x,2.1) {$\a$};
    \node at (\x,1.5) {$\b$};
    \node at (\x,1) {$\vdots$};
    \node at (\x,0.3) {$\c$};
  }
\end{scope}
\node at (6.1,3.35) {Codewords: $m\times\ell$};
\node[font=\figurefontsize] at (6.1,-0.35)
  {columns $C_1(x),\ldots,C_\ell(x)$};

\draw[arr] (8.05,1.2)--(10.25,1.2)
  node[midway,above,align=center,font=\figurefontsize]
  {sequential\\decoding};

\begin{scope}[shift={(10.4,0)}]
  \fill[black!12]  (0,0) rectangle (1.8,2.4);
  \fill[orange!25] (1.8,0) rectangle (2.4,2.4);
  \fill[white]     (2.4,0) rectangle (3.6,2.4);

  \draw (0,0) rectangle (3.6,2.4);
  \foreach \x in {0.6,1.2,1.8,2.4,3.0}
    \draw (\x,0)--(\x,2.4);
  \foreach \y in {0.6,1.2,1.8}
    \draw (0,\y)--(3.6,\y);

  \foreach \x/\a/\b/\c in {
    0.3/0/1/1,0.9/1/0/1,1.5/0/1/0} {
    \node at (\x,2.1) {$\a$};
    \node at (\x,1.5) {$\b$};
    \node at (\x,0.3) {$\c$};
  }
  \foreach \x in {0.3,0.9,1.5,2.1,2.7,3.3}
    \node at (\x,1) {$\vdots$};
  \foreach \y in {2.1,1.5,0.3}
    \node at (2.1,\y) {$?$};

  \draw[black,very thick] (1.8,0) rectangle (2.4,2.4);

  \draw[decorate,decoration={brace,amplitude=3pt}]
    (0,2.55)--(1.8,2.55)
    node[midway,above=5pt,font=\figurefontsize] {$1,\ldots,s-1$};
  \node[font=\figurefontsize] at (2.1,2.83) {$s$};
  \node[font=\figurefontsize] at (3,2.83) {unread};
\end{scope}
\node at (12.2,3.35) {Decoding: $m\times\ell$};
\node[font=\figurefontsize,align=center] at (11.15,-0.45)
  {recovered\\in every row};

\node[manager,text width=3.4cm] (mgr) at (12.2,-2.1)
  {\textbf{Manager}\\[2pt]
   $\operatorname{Dec}_1,\ldots,
     \operatorname{Dec}_s,\ldots$\\[2pt]
   one column at a time};

\draw[arr] (6.1,-0.7)--(6.1,-2.1)--(mgr.west)
  node[pos=0.48,above,align=center,font=\figurefontsize]
  {stored sketches\\$y_s=H_sC_s(x)$};

\coordinate (current) at (12.5,0);
\draw[arr] (mgr.north -| current)--(current)
  node[midway,right,align=left,font=\figurefontsize]
  {recover $C_s(x)$\\in all rows};
\end{tikzpicture}

\caption{Rows are encoded independently (left to middle).
The manager recovers columns sequentially (right), extending every
row's prefix by one bit at each step. The shaded prefix is recovered,
the outlined column is current, and the unfilled suffix is unrecovered.}
\label{fig:framework-matrices}
\end{figure}
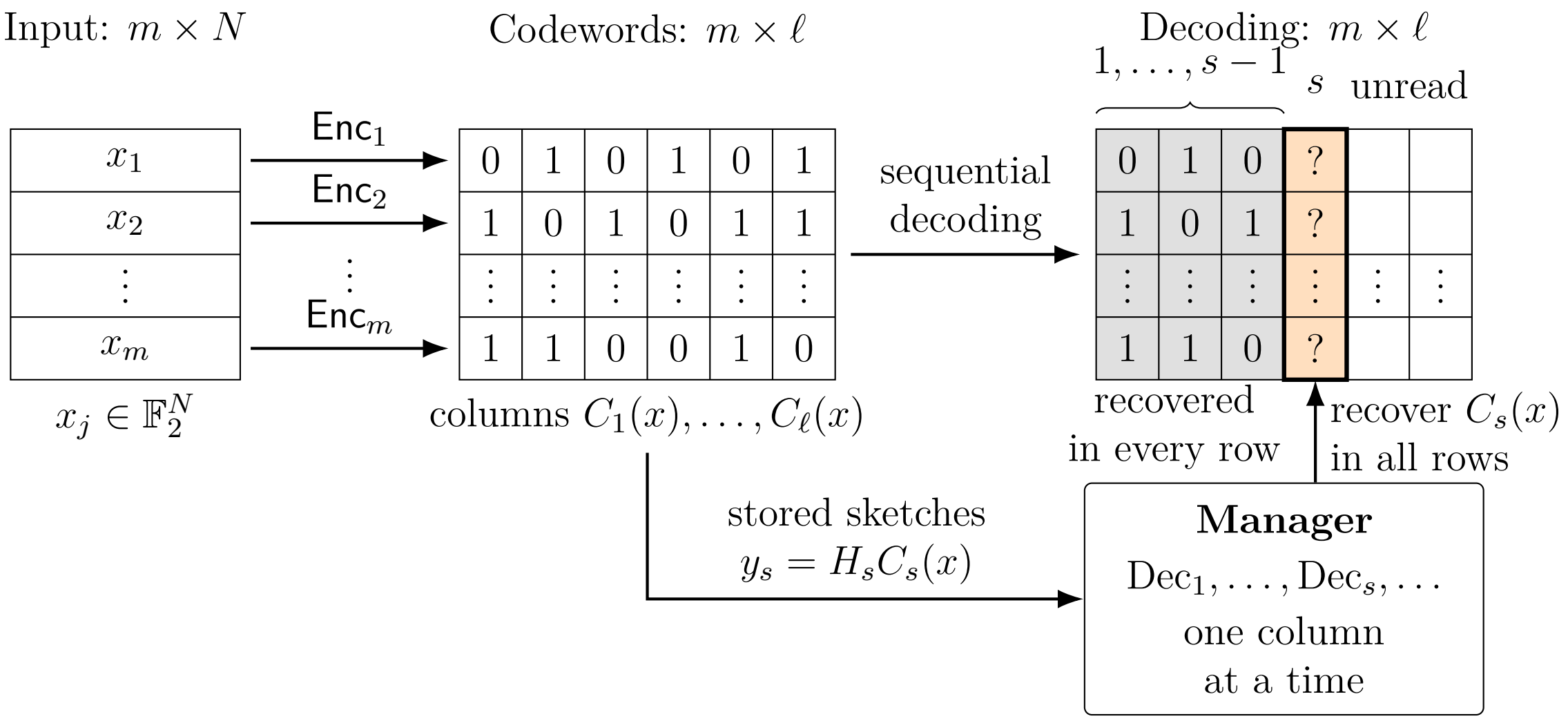

\paragraph{Column-by-column decoding framework.}
We now describe how hinted recovery is used to decode the measurements.
Arrange the input $x\in\mathbb F_2^n$ into an $m\times N$ matrix where $N=\ceil{n/m}$, padding with zeros
if necessary. Each row $x_j\in\F_2^N$ is encoded independently by
a binary linear map $E_j:\F_2^N\to\F_2^\ell$.
The resulting codewords form an $m\times\ell$ matrix, as illustrated
in Figure~\ref{fig:framework-matrices}.

Each row code supports \emph{decoding attempts from prefixes}.
Given the first $s-1$ recovered bits, its decoder may produce a
candidate for $x_j$ and use it to predict the next codeword bit.
If it has no candidate, it returns $\perp$.
A short prefix need not determine the row uniquely, so a prediction
may be incorrect. The concrete row codes and prediction rules will be specified in the next part.

We store each column $C_s(x)\in\F_2^m$ through a hinted recovery
sketch $H_sC_s(x)$ with public parameters $(m,X_s,Y_s)$, where $s\in [\ell]$.
To recover the columns in order, a \emph{manager} collects predictions
from the row decoders, viewed as \emph{workers}.
Let $J_s$ contain the workers returning $\perp$, and let $P_s$ contain
the predicted bits, with zeros on $J_s$.
The difference $C_s(x)-P_s$ is nonzero outside $J_s$ exactly where
a prediction is incorrect. By linearity, its sketch is
$H_s(C_s(x)-P_s)$.

The manager therefore invokes hinted recovery on this difference,
adds back $P_s$, and sends the recovered column $\widehat{C_s}(x)$ to the workers.
Their prefixes are now one bit longer, allowing the next round
of predictions. This step succeeds whenever there are at most
$X_s$ unspecified predictions and at most $Y_s$ incorrect predictions,
and the child decoder succeeds.

\paragraph{Residual recovery.}
After recovering all columns, the row decoders together produce a preliminary
estimate $\widehat x$ of the input.
The encoder also stores a CR sketch $\mathsf{CR}x$ for residual recovery.
By linearity, the decoder computes $\mathsf{CR}_{n,t}(x-\widehat x)=\mathsf{CR}_{n,t}x-\mathsf{CR}_{n,t}\widehat x$
and recovers the residual $x-\widehat x$, provided its weight is within
the predetermined recovery capacity $t$. Adding this residual to $\widehat x$
recovers $x$.
\begin{algorithm}[htbp]
\caption{Overall decoding process}
\label{alg:column-decoding}
\begin{algorithmic}
\STATE \textbf{Input} Column sketches $y_s=H_sC_s(x)$ for
  $s\in[\ell]$, and a residual sketch $a=\mathsf{CR}_{n,t}x$.
\STATE Initialize every row's recovered prefix as empty.
\FOR{$s=1,\ldots,\ell$}
  \STATE Each row decoder $j$ uses its prefix to produce
    $q_{j,s}\in\{0,1,\perp\}$.
  \STATE $J_s\gets\{j:q_{j,s}=\perp\}$.
  \STATE Form $P_s$ from the predictions, replacing $\perp$ by $0$.
  \STATE $\widehat D_s\gets\operatorname{Dec}_s(y_s+H_sP_s,J_s)$.
  \STATE $\widehat C_s\gets P_s+\widehat D_s$.
  \STATE Send $(\widehat C_s)_j$ to each row decoder $j$
    and append it to its recovered prefix.
\ENDFOR
\STATE Each row decoder performs final recovery using its full prefix
  to obtain an estimate $\widehat x_j$.
\STATE Assemble the row estimates into $\widehat x$ in the original
  coordinate order, discarding padding.
\STATE $\widehat e\gets\operatorname{Dec}_{\rm CR}(a+\mathsf{CR}_{n,t}\widehat x)$.
\STATE \textbf{return} $\widehat x+\widehat e$.
\end{algorithmic}
\end{algorithm}

\paragraph{Recursive structure.}
So far, we have described a single sketch instance, treating its column
sketches as subproblems. We now explain how these instances fit together
to solve the original sparse recovery problem for $x\in\F_2^n$.

We introduce two sketches: the \emph{top-level sketch} in
Section~\ref{sec:construction-top} and \HintSketch\ in
Section~\ref{sec:hintsketch-structure}. Both follow the framework above,
but use different row codes and corresponding prediction rules.
For the original input, we apply the top-level sketch, which recursively
invokes \HintSketch\ on each column. Each recursive \HintSketch\ instance
in turn invokes \HintSketch\ on its own columns, until the stopping
conditions in Section~\ref{sec:construction-parameters} are met.
Thus, the top-level construction is used only once; all subsequent
recursion takes place within \HintSketch.

\subsection{The top-level construction}
\label{sec:construction-top}

We instantiate the framework of
Section~\ref{sec:construction-framework} for the original input
$x\in\F_2^n$, using Pinsketch~\cite{DORS08} as the row code.

\paragraph{Encoding.}
Set $m=d$. Choose a public uniform permutation of $[n]$ and partition
the permuted coordinates into $d$ balanced rows of lengths
$N_1,\ldots,N_d$, padding each row to length $N=\ceil{n/d}$.
Put $w=\ceil{\log(N+1)}$.
Fix a basis of $\F_{2^w}$ and an embedding
$\iota:[N]\hookrightarrow\F_{2^w}^*$
as in Proposition~\ref{prop:pinsketch}.

Let $S_j\subseteq[N_j]$ be the support of row $j$.
We use $B=\min\{N,\ceil{C'\log(\log n+2)}\}$ words per row,
where $C'$ is a sufficiently large absolute constant.
For each $j\in[d]$ and $a\in[B]$, choose an independent public
uniform scalar
$\alpha_{j,a}\in\F_{2^w}^*$ and form
\begin{equation}
W_{j,a}:=\alpha_{j,a}P_{2a-1}(S_j)
=\alpha_{j,a}\sum_{v\in S_j}\iota(v)^{2a-1}.
\label{eq:construction-top-words}
\end{equation}
Each field element is represented by $w$ bits, which we call
a \emph{word}. The resulting binary measurement matrix has $d$ rows
and $\ell=Bw$ columns.
For each word position $a\in[B]$ and bit position $k\in[w]$, let the corresponding column $C_{a,k}(x)=((W_{j,a})_k)_{j=1}^d\in\F_2^d$.
We store $H_{a,k}C_{a,k}(x)$, where $H_{a,k}$ is a \HintSketch\
matrix with parameters $(d,X_{a,k},Y_{a,k})$.
Including the CR sketch for residual recovery, the stored vector is
\begin{equation}
H_rx=
\left(
  \bigl(H_{a,k}C_{a,k}(x)\bigr)_{a\in[B],\,k\in[w]},
  \CR_{n,t}x
\right).
\label{eq:construction-top-sketch}
\end{equation}
The parameters $X_{a,k},Y_{a,k}$ and residual capacity $t$ are chosen independently
of the input, as specified in Section~\ref{sec:top-child-parameters}.

\paragraph{Decoding.}
We follow the manager-worker and residual recovery procedure of
Algorithm~\ref{alg:column-decoding}, processing words in order
and bits within each word from $1$ to $w$.
It remains to specify the row decoder, namely the workers' behaviors.

Before word $a$, worker $j$ removes the public scalars $\alpha_{j,\star}$ from the first $a-1$ recovered words and uses Pinsketch to seek a set
$S\subseteq[N_j]$ of size at most $a-1$ matching these words.
Denote the resulting candidate by $\operatorname{Cand}_j(a-1)$, or
$\perp$ if none is found; set $\operatorname{Cand}_j(0)=\varnothing$.

The worker uses this candidate to predict word $a$.
It returns $\perp$ if no candidate is available or after a prediction
disagrees with the manager's recovered bit, and retries at the next
word boundary. The random scalar thus makes the first incorrect predicted bit occur at a suitably random position within a word.
Algorithm~\ref{alg:construction-top-row} gives the full procedure.

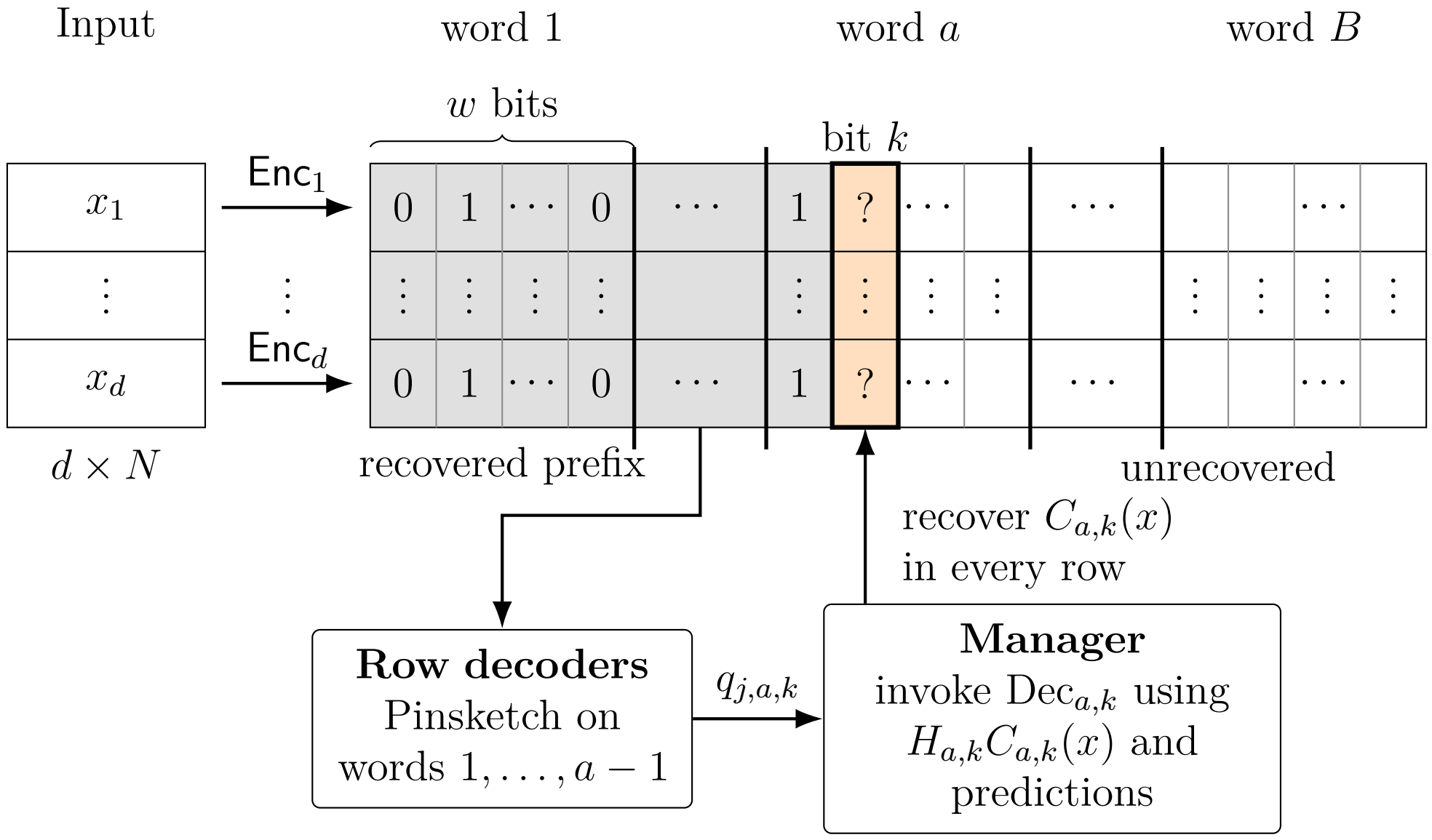
\begin{figure}[htbp]
\centering
\begin{tikzpicture}[
  font=\figurefontsize,
  arr/.style={-{Latex[length=2.5mm,width=1.8mm]},thick},
  box/.style={
    draw,rounded corners=2pt,align=center,inner sep=5pt
  },
  dots/.style={yshift=1mm}
]
\draw (0,0) rectangle (1.8,2.4);
\foreach \y in {0.8,1.6}
  \draw (0,\y)--(1.8,\y);
\node at (0.9,2) {$x_1$};
\node[dots] at (0.9,1.2) {$\vdots$};
\node at (0.9,0.4) {$x_d$};
\node at (0.9,3.65) {Input};
\node[font=\figurefontsize] at (0.9,-0.35) {$d\times N$};

\foreach \y/\j in {2/1,0.4/d}
  \draw[arr] (1.95,\y)--(3.15,\y)
    node[midway,above,font=\figurefontsize] {$\mathsf{Enc}_{\j}$};
\node[dots] at (2.55,1.2) {$\vdots$};

\fill[black!12]  (3.3,0) rectangle (7.5,2.4);
\fill[orange!25] (7.5,0) rectangle (8.1,2.4);
\draw (3.3,0) rectangle (12.9,2.4);
\foreach \y in {0.8,1.6}
  \draw (3.3,\y)--(12.9,\y);

\foreach \start in {3.3,6.9,10.5} {
  \begin{scope}[xshift=\start cm]
    \foreach \x in {0.6,1.2,1.8}
      \draw[black!45] (\x,0)--(\x,2.4);
    \foreach \x in {0.3,0.9,1.5,2.1}
      \node[dots] at (\x,1.2) {$\vdots$};
  \end{scope}
}

\foreach \x in {5.7,6.9,9.3,10.5}
  \draw[black,line width=1pt] (\x,-0.2)--(\x,2.55);

\foreach \y in {0.4,2} {
  \node at (3.6,\y) {$0$};
  \node at (4.2,\y) {$1$};
  \node at (4.8,\y) {$\cdots$};
  \node at (5.4,\y) {$0$};
  \node at (6.3,\y) {$\cdots$};
  \node at (7.2,\y) {$1$};
  \node at (7.8,\y) {$?$};
  \node at (8.4,\y) {$\cdots$};
  \node at (9.9,\y) {$\cdots$};
  \node at (12,\y) {$\cdots$};
}

\node at (4.5,3.65) {word $1$};
\node at (8.1,3.65) {word $a$};
\node at (11.7,3.65) {word $B$};
\node[font=\figurefontsize] at (7.8,2.65) {bit $k$};

\draw[very thick] (7.5,0) rectangle (8.1,2.4);
\draw[decorate,decoration={brace,amplitude=3pt}]
  (3.3,2.55)--(5.7,2.55)
  node[midway,above=4pt,font=\figurefontsize] {$w$ bits};

\node[font=\figurefontsize] at (4.5,-0.35) {recovered prefix};
\node[font=\figurefontsize] at (11.1,-0.35) {unrecovered};

\node[box,text width=3.1cm] (workers) at (4.5,-2.65)
  {\textbf{Row decoders}\\Pinsketch on words $1,\ldots,a-1$};
\draw[arr] (6.3,0)--(6.3,-0.8)-|(workers.north);

\node[box,text width=3.8cm] (mgr) at (9.5,-2.65)
  {\textbf{Manager}\\
   invoke $\operatorname{Dec}_{a,k}$ using\\
   $H_{a,k}C_{a,k}(x)$ and predictions};

\draw[arr] (workers.east)--(mgr.west)
  node[midway,above,font=\figurefontsize] {$q_{j,a,k}$};

\coordinate (current) at (7.8,0);
\draw[arr] (mgr.north -| current)--(current);
\node[anchor=west,align=left,font=\figurefontsize] at (8,-1.05)
  {recover $C_{a,k}(x)$\\in every row};
\end{tikzpicture}

\caption{Top-level encoding and decoding.
Each row is encoded into $B$ words of $w$ bits each.
The shaded region is the recovered prefix; the outlined column
is bit $k$ of word $a$.
Workers obtain candidates at word boundaries, and the manager
recovers one bit column at a time.}
\label{fig:construction-top-words}
\end{figure}

\begin{algorithm}[htbp]
\caption{The row decoder for the top-level sketch}
\label{alg:construction-top-row}
\begin{algorithmic}
\STATE \textbf{Input} Row index $j$ and access to the bits recovered
  by the manager.
\STATE Initialize the recovered prefix as empty.
\FOR{$a=1,\ldots,B$}
  \STATE $C\gets\operatorname{Cand}_j(a-1)$.
  \FOR{$k=1,\ldots,w$}
    \STATE $q_{j,a,k}\gets
  \bigl(\alpha_{j,a}P_{2a-1}(C)\bigr)_k$
  if $C\ne\perp$, and $\perp$ otherwise.
    \STATE Send $q_{j,a,k}$ to the manager and receive
      the recovered bit $v_{j,a,k}$.
    \STATE Append $v_{j,a,k}$ to the recovered prefix.
    \IF{$q_{j,a,k}\ne\perp$ and $q_{j,a,k}\ne v_{j,a,k}$}
      \STATE $C\gets\perp$.
    \ENDIF
  \ENDFOR
\ENDFOR
\STATE $C\gets\operatorname{Cand}_j(B)$.
\STATE \textbf{return} $C$ if $C\ne\perp$, and $\varnothing$ otherwise.
\end{algorithmic}
\end{algorithm}

\subsection{The {\mdseries\HintSketch} construction}
\label{sec:hintsketch-structure}

We now instantiate the framework of
Section~\ref{sec:construction-framework} for a hinted instance with
parameters $(N,X,Y)$ as in Problem~\ref{prob:hinted-recovery}.

All the parameters mentioned below, including $m,\ell,B_X,B_Y,c,t$ and the child parameters, will be specified in Section~\ref{sec:construction-parameters}.

\paragraph{Encoding.}
Roughly speaking, the encoding follows the same structure as the top-level
construction.
We first apply a public uniform permutation to $[N]$ and partition the coordinates into $m$ balanced rows.
The only difference is the encoding within each row. Instead of the
Pinsketch-based encoding used at the top level, we use a \emph{random binary
linear code}.

Let $S_j$ denote the row-local image of $\supp(x)$. For each $j\in[m]$, $v\in[N_j]$, and $s\in[\ell]$, choose an
independent public uniform bit $\beta_{j,v,s}\in\{0,1\}$, and define 
\begin{equation}
B_{j,s}(C):=\sum_{v\in C}\beta_{j,v,s},\quad C\subseteq [N_j].
\label{eq:hint-row-measurements}
\end{equation}
Thus the row encodings form an $m\times\ell$ matrix $(B_{j,s}(S_j))_{j\in[m],s\in[\ell]}$, where for each $s\in[\ell]$, its $s$th column is
$C_s(x):=
\bigl(B_{1,s}(S_1),\ldots,B_{m,s}(S_m)\bigr)\in\F_2^m$.
At a recursive node, each column is encoded by an independent child
{\HintSketch} $H_s$. Together with the residual sketch, we store
$\left(\bigl(H_sC_s(x)\bigr)_{s\in[\ell]},\;\CR_{N,t}x\right)$, where $H_s$ has parameters $(m,X_s,Y_s)$.

\paragraph{Decoding.}

We now describe the row decoder.
Let $J_j$ denote the row-local images of the hint $J$.
Intuitively, the row decoder first \emph{quotients out} the subspace generated by the vectors of $J_j$. It then searches, in increasing outside-hint support size, for a support whose code is consistent with the recovered prefix.
Once this explanation becomes unique, the row uses it to predict the next bit; otherwise it outputs $\perp$.

To be clear, suppose worker $j$ has recovered a prefix
$z_j^{(a)}\in\F_2^a$ of its row transcript. For $C\subseteq[N_j]$, write
$B_j^{(a)}(C):=
\bigl(B_{j,1}(C),\ldots,B_{j,a}(C)\bigr)$.
For an outside-hint support size $r$, define
$$
\mathcal C_{j,r}^{(a)}
:=
\left\{
Z\cup A:
Z\subseteq J_j,\;
A\subseteq[N_j]\setminus J_j,\;
|A|=r,\;
B_j^{(a)}(Z\cup A)=z_j^{(a)}
\right\}
$$
as the family of all row supports that are
consistent with the recovered prefix, where exactly $r$ support coordinates lie outside the hint $J_j$.

For each fixed $A$, the possible sets $Z\subseteq J_j$ are obtained
by solving a binary linear system. Enumerating $A\subseteq [N_j]\setminus J_j$ therefore lets the
worker determine whether $\mathcal C_{j,r}^{(a)}$ is empty, a singleton,
or contains multiple supports, without enumerating the subsets of $J_j$.
To avoid predicting from an insufficient prefix, define
\begin{equation}
\sigma_{j,r}:=
\ceil{\log\left(
2^{|J_j|}\sum_{h=0}^{r}\binom{N_j-|J_j|}{h}
\right)}.
\label{eq:hint-prefix-condition}
\end{equation}
The quantity inside the logarithm counts all row supports with at most
$r$ coordinates outside $J_j$. The worker tests size $r$ only after
recovering $\sigma_{j,r}+c$ bits.
Here $c$ is an extra parameter whose value is specified in Section~\ref{sec:construction-parameters}.
\footnote{Intuitively, there are at most $2^{\sigma_{j,r}}$ candidate supports of outside-hint size at most $r$, and each random measurement halves the probability that a fixed wrong candidate remains consistent. Waiting for $\sigma_{j,r}+c$ bits therefore reduces this ambiguity.}
To control the computational cost, we impose two bounds: $B_X$ on the hint size handled by each worker, and $B_Y$ on the outside-hint support size it enumerates. Both are specified in Section~\ref{sec:construction-parameters}.
\begin{algorithm}[htbp]
\caption{The row decoder for \HintSketch}
\label{alg:hint-row}
\begin{algorithmic}
\STATE \textbf{Input} Row index $j$, hint $J_j$, and access to the
  manager's recovered bits.
\IF{$|J_j|>B_X$}
    \STATE Instruct the manager to use $\perp$ for this row at every position.
  \STATE \textbf{return} $\varnothing$.
\ENDIF
\STATE Initialize $z_j^{(0)}$ as the empty prefix.
\FOR{$a=0,\ldots,\ell-1$}
  \STATE $C\gets\perp$.  
\FOR{$r=0,\ldots,\min\{B_Y,N_j-|J_j|\}$}
  \IF{$a\ge\sigma_{j,r}+c$}
    \STATE Determine whether $\mathcal C_{j,r}^{(a)}$
      is a singleton by enumeration and linear algebra.
    \IF{$\mathcal C_{j,r}^{(a)}=\{C'\}$}
      \STATE $C\gets C'$; \textbf{break}.
    \ENDIF
  \ENDIF
\ENDFOR
  \STATE $q_{j,a+1}\gets B_{j,a+1}(C)$ if $C\ne\perp$,
    and $\perp$ otherwise.
  \STATE Send $q_{j,a+1}$ to the manager and receive $v_{j,a+1}$.
  \STATE $z_j^{(a+1)}\gets(z_j^{(a)},v_{j,a+1})$.
\ENDFOR
\STATE Determine whether
  $\displaystyle\bigcup_{0\le r\le \min\{B_Y,N_j-|J_j|\}}
    \mathcal C_{j,r}^{(\ell)}$
  has a unique member $C$.
\STATE \textbf{return} $C$ if unique, and $\varnothing$ otherwise.
\end{algorithmic}
\end{algorithm}

\paragraph{Stopping the recursion.}
The parameter rules in Section~\ref{sec:construction-parameters}
may try to stop the recursion at a node. In this case, we use one of two constructions:
1) \textbf{Direct:} store all $\ell$ columns directly,
together with $\CR_{N,t}x$, and apply final row recovery and residual
correction;
2) \textbf{CR-only:}  store only $\CR_{N,D}x$, where $D=\min\{N,X+Y\}$,
and recover $x$ directly.
The choice is specified in Section~\ref{sec:construction-parameters}.

%% file: 2609-version/4prob-tools.tex
\section{Probabilistic tools and the decoding framework}
\label{sec:analysis-framework}

We give the probability bounds and the framework used to analyze
both constructions.

\subsection{Concentration and entropy bounds}
\label{sec:analysis-probability}

We first prepare several probabilistic tools which will be used below.
Proofs of the lemmas in this subsection are given in
Appendix~\ref{app:probabilistic-tools}.

\begin{lemma}[Chernoff bound for hypergeometric random variables]
\label{lem:balanced-load}
Let $S$ be a uniformly random $K$-subset of $[N]$, and let
$Z=|S\cap I|$ for a fixed $q$-subset $I\subseteq[N]$.
Writing $\mu=Kq/N>0$, we have
\begin{equation}
\forall t\ge0,\quad
\Pr[Z\ge\mu+t]\le
\exp\!\left(-\frac{t^2}{2(\mu+t)}\right).
\label{eq:hypergeometric-tail}
\end{equation}
In particular, for $k\ge2\mu$, $\Pr[Z\ge k]\le e^{-k/8}$.
\end{lemma}

We use a slight extension of McDiarmid's inequality that allows a uniform random injection in addition to independent random variables.
Intuitively, we will apply it by expressing each statistic as a function of all the randomness involved and verifying the bounded-differences conditions to give a concentration bound.

\begin{lemma}[An extension of McDiarmid's inequality]
\label{lem:injection-concentration}
Let $\pi:[D]\hookrightarrow[N]$ be a uniform injection, and let
$R=(R_1,\ldots,R_b)$ be random variables such that $\pi,R_1,\ldots,R_b$ are mutually independent.

Let $F(\pi,R)$ be a real-valued function. Suppose there exist constants $L_p,L_r\ge0$ such that:
\begin{itemize}
  \item For every $\pi,R$ and every transposition $\tau$ of $[N]$,
    $|F(\tau\circ\pi,R)-F(\pi,R)|\le L_p$.
  \item For every $\pi$ and every $R,R'$ differing in one coordinate,
    $|F(\pi,R')-F(\pi,R)|\le L_r$.
\end{itemize}
If $DL_p^2+bL_r^2>0$, then for every $t>0$,
\begin{equation}
\Pr[F-\mathbb EF\ge t]
\le \exp\left(-\frac{2t^2}{DL_p^2+bL_r^2}\right).
\label{eq:injection-concentration}
\end{equation}
\end{lemma}

To analyze row decoding, we also bound the probability that an
incorrect support matches the true measurements.

\begin{lemma}[Separation by random binary measurements]
\label{lem:random-separation}
For each $v\in[N]$, independently choose a uniform vector
$\beta(v)\in\F_2^q$, and define
$B(C)=\sum_{v\in C}\beta(v)$ for $C\subseteq[N]$.
For any fixed support $S\subseteq[N]$ and fixed family
$\mathcal A\subseteq 2^{[N]}$,
\begin{equation}
\Pr\bigl[\exists C\in\mathcal A\setminus\{S\}: B(C)=B(S)\bigr]
\le |\mathcal A\setminus\{S\}|\,2^{-q}.
\label{eq:random-separation}
\end{equation}
\end{lemma}

The { following} lemmas let us sum the space required by the children:
the entropy identity handles the outside-hint cost, and the
prefix-tail bounds control the hint capacities.

\begin{lemma}[Entropy and total mass]
\label{lem:entropy-mass}
For a finite family of nonnegative $q_i$ with $0<Q=\sum_iq_i\le N$,
\begin{equation}
\sum_i h_N(q_i)=h_N(Q)+Q\mathsf H((q_i/Q)_i).
\label{eq:entropy-mass}
\end{equation}

In particular, $h_N(u+v)\le h_N(u)+h_N(v)$ for $u,v\ge0$ with $u+v\le N$.

\end{lemma}

\begin{lemma}[{Entropy from the mean}]
\label{lem:entropy-mean}
If a random variable $T$ takes values in the nonnegative integers and has finite mean,
then $\mathsf H(T)\le\log(e(1+\mathbb ET))$.
\end{lemma}

\begin{lemma}[Summing prefix tails]
\label{lem:prefix-sums}
For $z\ge0$ and integers $a\ge0$, put
$G_a(z)=\min\{1,2^{z-a}\}$.
There is an absolute constant $C$ such that
\begin{equation}
\sum_{a\ge0}G_a(z)\le z+C,
\qquad
\sum_{a\ge0}aG_a(z)\le C(1+z)^2.
\label{eq:prefix-sums}
\end{equation}
\end{lemma}

\subsection{From row estimates to recovery}
\label{sec:analysis-exact-columns}
\label{sec:analysis-composition}

We first analyze a run in which the manager returns the true column bits to the workers after each prediction (\textit{i.e.}, the subroutines return correct answers).
For each column, we count workers returning $\perp$ and workers predicting the wrong bit.
After the final row recovery, we also count the remaining errors which will be handled by CR Sketch.
We then prove that the decoder succeeds whenever these counts fit the chosen capacities.

\begin{definition}[Execution with exact child answers]
\label{def:exact-child}
Fix a node's input $x$ and hint $J$.\footnote{The top layer has no input hint, and we set $J=\varnothing$ there so that
all definitions and bounds below apply to both top-layer and HintSketch nodes.}
Run the row decoders as in Algorithm~\ref{alg:column-decoding}, giving each worker its bit of the true column $C_s(x)$ after it predicts column $s$.
Let $U_s$ count workers returning $\perp$ and $V_s$ count incorrect predictions in column $s$.
Let $Z=\wt{x+\widehat x}$ be the residual weight after final row recovery.
\end{definition}

The quantities $U_s$ and $V_s$ are exactly the hint and outside-hint
capacities needed by the child for column $s$.
In fact, $U_s\le X_s$ and $V_s\le Y_s$ are precisely the child's two
input conditions in Problem~\ref{prob:hinted-recovery}.
(To see why, consider that the input of the child is $C_s(x)+P_s$ and its hint $J_s$ contains exactly the rows returning $\perp$, so $|J_s|=U_s$ and $|\supp(C_s(x)+P_s)\setminus J_s|=V_s$.)
Similarly, if $Z\le t$, the residual has at most $t$ errors, which the final CR sketch can correct.
We next show that $U_s,V_s,Z$ concentrate around their expectations by applying Lemma~\ref{lem:injection-concentration}.

\begin{lemma}[Concentration]
\label{lem:public-capacities}
Fix a node's input $x$ and hint $J$.
Let $m$ be its number of rows and $D\ge m$ an upper bound on $|J\cup\supp(x)|$.
Suppose each final row estimate $\widehat x_j$ has at most $K\ge1$ nonzero coordinates. For every $\Delta>0$,
\begin{equation}
\begin{aligned}
\Pr[U_s>\mathbb EU_s+\Delta]&\le \exp\!\left(-\frac{2\Delta^2}{5D}\right),\qquad \forall s\in[\ell],\\
\Pr[V_s>\mathbb EV_s+\Delta]&\le \exp\!\left(-\frac{2\Delta^2}{5D}\right),\qquad \forall s\in[\ell],\\
\Pr[Z>\mathbb EZ+\Delta]&\le \exp\!\left(-\frac{\Delta^2}{10DK^2}\right).
\end{aligned}
\label{eq:common-capacity-tails}
\end{equation}
\end{lemma}

\begin{proof}
Apply Lemma~\ref{lem:injection-concentration} to the placement of $J\cup\supp(x)$ and the independent randomness of the $m$ rows.
A position swap of $J\cup\supp(x)$ affects at most two rows, changing $U_s,V_s$ by at
most $2$.
Replacing one row's randomness changes each count by at
most $1$.
Since $m\le D$, the concentration denominator is at most
$4D+m\le5D$.

For the bound of $Z$, we denote $S_j=\supp(x_j)$ and $\widehat S_j=\supp(\widehat x_j)$ in row coordinates, then
\[
Z=\sum_{j=1}^m|S_j\triangle\widehat S_j|
 =\wt{x}+\sum_{j=1}^m\bigl(|\widehat S_j|-2|S_j\cap\widehat S_j|\bigr).
\]
Each term in the last sum lies in $[-K,K]$ and depends only on the
placement of $J\cup\supp(x)$ in row $j$ and that row's randomness.
The sensitivities are therefore $4K$ for a position swap and $2K$ for a change to one row's randomness.
The denominator is at most $16DK^2+4mK^2\le20DK^2$, giving the last bound.
\end{proof}

While Lemma~\ref{lem:public-capacities} gives the single-node-error bounds, Theorem~\ref{thm:decoder-composition} combines them by a union bound to give the failure probability for the entire tree.

\begin{theorem}[Recovery from capacity bounds]
\label{thm:decoder-composition}
Fix a valid root input $x$ (together with its hint if present) satisfying $\|x\|_0\le d$.
If $U_s\le X_s$, $V_s\le Y_s$, and $Z\le t$ hold at every node, the decoder always recovers the input without error.
If each of the above bounds fails with probability at most $\delta$, the total failure probability is at most $Q\delta$, where $Q$ is the total number of these bounds in the tree.
\end{theorem}

\begin{proof}
We prove by induction that, if all capacity bounds in a node's subtree hold, that node returns the correct reply.
Obviously, the claim holds at leaves by direct storage and CR recovery.
At an internal node, process the columns in order. Given correct earlier columns, $U_s\le X_s$ and $V_s\le Y_s$ ensure that the next child's input satisfies its size bounds, so the induction hypothesis gives the correct column. After all columns are recovered, $Z\le t$ guarantees that CR repair recovers the current node's input.
For the latter claim, consider that each node's input and hint depend only on ancestor randomness.
If all ancestor bounds hold, each of the above bounds at any node fails with conditional probability at most $\delta$ over that node's independent internal randomness.
A union bound over the $Q$ possible first violated bounds along root-to-leaf paths gives $Q\delta$.
\end{proof}

For specific usage, we will add a sufficiently large margin to upper bounds on $\mathbb EU_s,\mathbb EV_s,\mathbb EZ$ to obtain capacities $X_s,Y_s,t$ such that $U_s\le X_s$, $V_s\le Y_s$ and $Z\le t$ hold simultaneously with high probability.
As the upper bounds depend only on the construction parameters, not on $x$ or $J$ (as we will discuss in Sections~\ref{sec:analysis-hint} and~\ref{sec:analysis-top}), both parties can therefore compute them and fix the same margin in advance.

%% file: 2609-version/5hintsketch.tex
\section{Performance of {\mdseries\HintSketch}}
\label{sec:analysis-hint}
\label{sec:construction-parameters}

This section proves the performance guarantee for a {\HintSketch} node in Theorem~\ref{thm:hint-forest}, which Section~\ref{sec:analysis-top} uses as a black box.

\subsection{Performance guarantee and proof overview}
\label{sec:hint-proof-overview}

The main result of this section is:

\begin{theorem}[Performance of {\mdseries\HintSketch}]
\label{thm:hint-forest}
Recall that $h_N(z)=z\log(eN/z)$.
For a hinted node with parameters $(N,X,Y)$, there are public parameter choices such that, for every fixed input and hint satisfying
\eqref{eq:hint-promise}, recovery fails with probability at most
$2^{-\lambda}$. The sketch has length at most
\begin{equation}
(1+O(1/\log n))\bigl(X+\alpha h_N(Y)\bigr)
+O(N/\log^8 n)+\lambda\log^{O(1)} n,
\label{eq:hint-forest-length}
\end{equation}
where $\alpha$ is an absolute constant.
Note that $N$ denotes the size of the current subproblem, while $n$ denotes the size of the original problem.
It is binary linear, with column locality and update time
$\log^{O(1)} n$, and worst-case decoding time
$N\exp(O((\log\log n)^2))$.
\end{theorem}

We prove the space bound by induction, taking $\alpha=C^2$ for a
sufficiently large constant $C$ and calling $X+C^2h_N(Y)$ the node's
\emph{charge}.
The inductive step makes the children's total charge at most the
parent's. Concretely, a sequence of lemmas establishes the following
bounds on the child capacities in Lemma~\ref{lem:hint-local-cost}:
\[
\sum_sX_s\le X+O(C h_N(Y)),\qquad
\sum_sh_Y(Y_s)\le\tfrac14h_N(Y).
\]
Thus the decrease in outside cost pays for the additional hints
(Figure~\ref{fig:hint-budget}). The remaining costs, including residual
repair, are absorbed into a small relative overhead on the charge
and the additive term in the theorem.

\input{2609-version/fig-hint-budget}

\edrev
\subsection{Row Predictions}
\label{sec:hint-row-parameters}
\label{sec:hint-stopping}
\label{sec:hint-row-analysis}
\label{sec:hint-residual}

\paragraph{Parameter choices and stopping rules.}
Put $D=\min\{N,X+Y\}$.
We choose enough rows in {\mdseries\HintSketch} to ensure that the expected outside-hint load for a single row is $O(1)$ and the expected hint load is $O(\log^3 n)$.
We also choose a small parameter $\eta$ to control the residual errors to $\mathbb E Z\le \eta(D+1)$ and aim for failure probability $2^{-\lambda_*}$ for each capacity bound mentioned in Theorem \ref{thm:decoder-composition}.
Formally, we first fix a sufficiently large absolute constant $A$, then choose
$C\ge A$ sufficiently large. Then set
\[
\begin{gathered}
m=\max\{Y,\ceil{X/\log^3 n}\},\quad \eta=\log^{-A}n,\\
\lambda_*=\lambda+C\ceil{\log\log(n+2)}+C,\quad
 t=\min(N,\lceil 2\eta(D+1)\rceil).
\end{gathered}
\]

For every recursive step, we use the first applicable rule:
\begin{enumerate}
\item \textbf{Cheap leaf}: if $D\le N/\log^{10}n$, use $\CR_{N,D}$.
\item \textbf{Small-instance leaf}: if $D\le C^3\lambda_*\log^6 n/\eta^2$,
      use $\CR_{N,D}$.
\item \textbf{Recursive node}: if $m=Y$, sketch the columns recursively.
\item \textbf{Direct node}: otherwise, store the columns directly.
\end{enumerate}

In either of the last two cases, each row contains at most
\begin{equation}
N_{\max}=\ceil{N/m}\le 1+(\log^3 n+1)\log^{10}n=O(\log^{13} n)
\label{eq:nmax-bound}
\end{equation}
coordinates, as $N<D\log^{10}n\le(X+Y)\log^{10}n\le m(\log^3n+1)\log^{10}n$ in these cases.
Using the same constant $C$, we choose\footnote{Here, unsubscripted $O$ and $\Omega$ below have constants independent of $A,C$. Recall that $C\ge A$.}
\begin{equation}
\begin{aligned}
B_X&=\min\{N_{\max},\ceil{\log^3 n+C\log^2 n}\},\\
B_Y&=\min\{N_{\max},\ceil{C\log(\log n+2)}+4\},\\
\ell&=B_X+B_Y\ceil{\log(eN_{\max})}
       +B_Y+\ceil{\log(16/\eta^2)}+10\\
&\le\log^3 n+O(C\log^2 n+A\log\log n)\le \log^3 n+O(C\log^2 n),
\end{aligned}
\label{eq:hint-cap-choices}
\end{equation}
where $\ell$ is the number of columns and $B_X,B_Y$ are bounds used in Algorithm \ref{alg:hint-row}.
Thus every row output has size at most $K:=B_X+B_Y=O(C\log^3 n)$.
Substituting $B_X,B_Y\le N_{\max}=O(\log^{13}n)$ into \eqref{eq:hint-cap-choices} also gives $\ell=O(A\log^{14}n)$, independently of $C$.

Also, a direct node stores
\begin{equation}
\label{eq:hint-direct-storage}
\begin{aligned}
m\ell&\le (X/\log^3 n+1)(\log^3 n+O(C\log^2 n))\\
&\le X+O(CX/\log n)+O(C\log^3 n)
\end{aligned}
\end{equation}
bits.

\paragraph{Error bounds.}
We now analyze the error bounds of the last two cases, as the former two cases are error-free.
Write $y_j=|S_j\setminus J_j|$.
We say a row is \textit{regular} if $|J_j|\le B_X$ and $y_j\le B_Y$. Otherwise, we say the row is \emph{overloaded}, and its worker may fail.
We next bound the probability that a row worker returns \(\perp\) or predicts the next bit incorrectly after recovering \(a\) measurements.
Using $G_a(z)=\min\{1,2^{z-a}\}$ from Lemma~\ref{lem:prefix-sums}, we obtain the following bounds.

\begin{lemma}
\label{lem:hint-prediction}
For a regular row, after $a$ recovered bits,
\begin{equation}
\Pr[\text{worker returns }\perp]\le G_a(\sigma_{j,y_j}+c),\qquad
\Pr[\text{incorrect prediction}]\le2^{-c}G_a(\sigma_{j,y_j}+c).
\label{eq:hint-row-prediction}
\end{equation}
Recall that $\sigma$ is defined at (\ref{eq:hint-prefix-condition}) and $c$ is defined after that.
\end{lemma}
\begin{proof}

Each incorrect support matches the first $a$ true measurements with probability $2^{-a}$ (Lemma~\ref{lem:random-separation}).
When $a\ge\sigma_{j,y_j}+c$, the true support $S_j$ is eligible.
Unless an incorrect candidate with at most $y_j$ outside-hint positions matches the prefix, the worker reaches size $y_j$ and uniquely identifies $S_j$.
Hence, both returning $\perp$ and predicting incorrectly have probability at most $2^{\sigma_{j,y_j}-a}$, which gives
$$
\Pr[\text{worker returns }\perp]
\le
\begin{cases}
1, & a<\sigma_{j,y_j}+c,\\
2^{\sigma_{j,y_j}-a}, & a\ge \sigma_{j,y_j}+c.
\end{cases}
$$
Thus, $\Pr[\text{worker returns }\perp]\le \min\{1,2^{\sigma_{j,y_j}+c-a}\}$.

For incorrect-prediction probability, we additionally analyze the case of $a<\sigma_{j,y_j}+c$.
In our algorithm, the worker tests only sizes $r$ such that $\sigma_{j,r}\le a-c$.
Since $2^{\sigma_{j,r}}$ bounds the number of supports with at most $r$ outside-hint coordinates, at most $2^{a-c}$ candidates are eligible in total.
A union bound gives $\Pr[\text{incorrect prediction}]\le 2^{-c}$ for $a<\sigma_{j,y_j}+c$. So
$$
\Pr[\text{incorrect prediction}]\le\min\{2^{-c},2^{\sigma_{j,y_j}-a}\}\le 2^{-c}\min\{1,2^{\sigma_{j,y_j}+c-a}\},
$$
which proves \eqref{eq:hint-row-prediction}.
\end{proof}

\begin{lemma}[Row recovery]
\label{lem:hint-final}
\label{lem:row-recovery-fail-bound}
For any regular row, final recovery for this row fails with probability at most $\eta^2/8$.
With these choices, the expected number of coordinates from $J\cup\supp(x)$ in overloaded rows is at most $D\eta^2$.
\end{lemma}

\begin{proof}
For a regular row, final recovery fails only if another valid support matches all $\ell$ measurements.
As there are at most $2^{B_X}(eN_{\max})^{B_Y}$ allowed supports,\footnote{Note that $\sum_{r=0}^{B_Y}\binom{N_{\max}}{r}\le (eN_{\max})^{B_Y}$.}
a union bound gives
\[
\Pr[\text{final recovery fails}]
\le 2^{B_X+B_Y\log(eN_{\max})-\ell}\le\eta^2/8.
\]
For the last claim, fix a coordinate in $J\cup\supp(x)$ and condition on its row.
The remaining hint and outside-hint loads of this row are hypergeometric, with means at most $X/m+O(1)\le\log^3n+O(1)$ and $Y/m+O(1)=O(1)$.
The fixed coordinate adds at most one to either load.

If $B_X$ or $B_Y$ is capped at $N_{\max}$, the corresponding condition must hold.
Otherwise, the thresholds $B_X,B_Y$ exceed these means by order $C\log^2n$ and $C\log(\log n+2)$, respectively.
Lemma~\ref{lem:balanced-load} gives
\[
\begin{aligned}
\Pr[\text{hint load}>B_X]
&\le \exp\left(-\dfrac{C^2\log^2 n}{2(C+\log n)}\right)\le e^{-\Omega(C\log n)},\\
\Pr[\text{outside-hint load}>B_Y]
&\le e^{-B_Y/8}\le e^{-\Omega(C\log(\log n+2))}.
\end{aligned}
\]
Since $\eta^2=(\log n)^{-2A}$, choosing $C$ as a sufficiently large
constant multiple of $A$ makes each probability at most $\eta^2/2$.
A union bound thus shows that the probability of overload is at most $\eta^2$.
By linearity of expectation, the expected number of such coordinates
is at most $|J\cup\supp(x)|\eta^2\le D\eta^2$.
\end{proof}

\begin{lemma}[Remaining errors]
We have $\mathbb EZ\le\eta(D+1)$.
For $t=\min(N,\ceil{2\eta(D+1)})$, $\Pr[Z>t]\le2^{-\lambda_*}$.
\end{lemma}

\begin{proof}
Each row contributes at most $|S_j|+K$ errors as the decoder outputs at most $K$ positions and its symmetric difference with $S_j$ has size at most $|S_j|+K$.
For overloaded rows,
\[
\sum_{j:\,\mathrm{overloaded}}(|S_j|+K)
\le (K+1)\sum_{j:\,\mathrm{overloaded}}|J_j\cup S_j|.
\]
Taking expectations gives a contribution of at most $(K+1)D\eta^2$ by Lemma \ref{lem:row-recovery-fail-bound}.

For regular rows, $|S_j|,|\widehat S_j|\le K$, and
$\widehat S_j=S_j$ whenever $S_j=\varnothing$: then $\varnothing$ is consistent with the full recovered transcript, so the decoder outputs $\varnothing$ whether or not the consistent support is unique.
By Lemma~\ref{lem:row-recovery-fail-bound}, each nonempty regular row contributes
at most $2K\Pr[\widehat S_j\ne S_j]\le2K\eta^2/8$ expected errors.
There are at most $D$ such rows, and empty supports recover exactly.
Thus
\[
\mathbb EZ\le(K+1)D\eta^2+2KD\eta^2/8\le\eta(D+1),
\]
using $K=B_X+B_Y\le 2N_{\max}=O(\log^{13}n)$ and taking $A$ sufficiently large.

For $t<N$, Lemma~\ref{lem:public-capacities} gives
\[
\Pr[Z>t]\le
\exp\left(-\frac{\eta^2(D+1)^2}{10DK^2}\right)
\le \exp\left(-\Omega\left(\frac{\eta^2D}{C^2\log^6n}\right)\right)
\le \exp(-\Omega(C\lambda_*))
\le 2^{-\lambda_*},
\]
using $K=O(C\log^3n)$,
$D>C^3\lambda_*\log^6n/\eta^2$ as the second stopping rule does not apply, and sufficiently large $C$.
For $t=N$, the probability is zero.
\end{proof}

\subsection{Child capacities and space contraction}
\label{sec:hint-capacities}
\label{sec:hint-mean-bounds}
\paragraph{Choosing child capacities.}
We now choose $X_s,Y_s$ from the parameters so that $U_s\le X_s$ and $V_s\le Y_s$ hold with high probability.
As discussed in the final part of Section \ref{sec:analysis-composition}, we do this by bounding their expectations and adding a margin.

At a recursive node, there must be $m=Y$. Set $c=2\ceil{\log(N/Y)}+C$. At a direct leaf, take $c=0$.
To obtain bounds independent of the actual input $x,J$, consider a reference placement as follows.

\begin{definition}[Reference Placement]
Distribute $D-Y$ hint labels and $Y$ outside-hint nonzero labels uniformly at distinct positions in $[N]$.
Let $J_j^*$ be the hint set in row $j$ in this reference placement,
and let $y_j^*$ count its outside-hint nonzero positions, as they correspond to $J_j$ and $y_j$ for the actual input.    
\end{definition}
Their joint distribution depends only on $N,D,Y$ and the row sizes, not on $x,J$.
Define
\[
\sigma_j^*=\ceil{\log\left(
2^{|J_j^*|}\sum_{r=0}^{y_j^*}\binom{N_j-|J_j^*|}{r}\right)},\qquad
Q_s=\sum_j\mathbb E G_{s-1}(\sigma_j^*+c),\quad 1\le s\le\ell.
\]
For each column $s$, choose
\begin{equation}
X_s=\min\{Y,\ceil{Q_s+2\eta(D+1)}\},\qquad
Y_s=\min\{Y,\ceil{2^{-c}Q_s+2\eta(D+1)}\}.
\label{eq:hint-public-children}
\end{equation}
These quantities depend only on the parameters, so both parties can
compute the same capacities.

\paragraph{Verifying the capacities and bounding space.}
We now prove that the capacities are sufficient.
\begin{lemma}[Sufficient child capacities]
\label{lem:hint-child-capacities}
At a recursive node, for every input satisfying \eqref{eq:hint-promise}
and every column $1\le s\le\ell$,
\[
\mathbb E U_s\le Q_s+D\eta^2,\qquad
\mathbb E V_s\le2^{-c}Q_s+D\eta^2.
\]
With the capacities in \eqref{eq:hint-public-children}, each of the events
$U_s>X_s$ and $V_s>Y_s$ has probability at most $2^{-\lambda_*}$.
\end{lemma}
\begin{proof}
We prove this by comparing the actual input with the reference placement.
Arbitrarily add hint positions until $|J\cup\supp(x)|=D$, then reclassify hint positions as outside-hint nonzeros until there are $Y$.
As both operations enlarge the candidate family, applying the same random permutation to the original and enlarged sets therefore gives $\sigma_{j,y_j}\le\sigma_j^*$.
Lemmas~\ref{lem:hint-prediction} and~\ref{lem:hint-final} then give
\[
\begin{aligned}
\mathbb E U_s&\le\sum_{j=1}^{m}(\mathbb E[G_{s-1}(\sigma_{j,y_j}+c)]+\operatorname{Pr}[j\text{ is overloaded}])\\
&\le Q_s+D\eta^2\le Q_s+\eta(D+1),\\
\mathbb E V_s&\le2^{-c}Q_s+D\eta^2\le2^{-c}Q_s+\eta(D+1).
\end{aligned}
\]
Since $U_s,V_s\le Y$, Lemma~\ref{lem:public-capacities} yields $\Pr[U_s>X_s]
\le\Pr[U_s>\mathbb E U_s+\eta(D+1)]\le2^{-\lambda_*}$. The same argument applies to $V_s$.
\end{proof}

\begin{definition}[Density]
At a node $u$ with parameters $(N,X,Y)$, define its density as $\rho:=Y/N$.
Obviously, there must be $0\le \rho\le 1$.
\end{definition}
\begin{lemma}
\label{lem:hint-entropy-cost}
At a recursive node, for sufficiently large $C$, $\sum_{s=1}^{\ell}h_Y(2^{-c}Q_s)\le h_N(Y)/8$.
\end{lemma}
\begin{proof}
Using $\sigma_j^*\le |J_j^*|+y_j^*\log(eN_{\max})+1$ and $\sum_j |J^*_j|\le X,\sum_j y^*_j=Y$,
Lemma~\ref{lem:prefix-sums} gives (note that $Q_1\ge Y$)
\begin{equation}
\label{eq:lemma-5-8-bound-of-Qs}
\begin{aligned}
    Y\le\sum_sQ_s
&=\mathbb E\left[\sum_s\sum_jG_{s-1}(\sigma_j^*+c)\right]\le\mathbb E\left[\sum_j\sigma_j^*\right]+O(cY)\\
&\le X+Y\log(eN_{\max})+O(cY)
\le X+O(C h_N(Y)),
\end{aligned}
\end{equation}
where the last step uses $N_{\max}= \lceil N/Y\rceil\le2/\rho$, and $c\le C+2\log(1/\rho)+2$.

For each $j$, the counts $|J_j^*|$ and $y_j^*$ are hypergeometric, with means at most $2X/Y$ and $2$, respectively, and variances at most their means. Thus, $\mathbb E(|J_j^*|^2)\le \mathbb E|J_j^*|(1+\mathbb E|J_j^*|)\le (2X/Y+1)\mathbb E|J_j^*|\implies \sum_j\mathbb E(|J_j^*|^2)\le (2X/Y+1)(D-Y)\le X(2X/Y+1)$ and $\mathbb E {y_j^*}^2=O(1)$.
Lemma~\ref{lem:prefix-sums} gives

\begin{equation}
\begin{aligned}
\sum_s(s-1)Q_s
&\le O\!\left(\sum_j\mathbb E(1+\sigma_j^*+c)^2\right)\\
&\le O\!\left(\sum_j\mathbb E(|J_j^*|+y_j^*\log(eN_{\max})+c+2)^2\right)\\
&\le O\!\left(\sum_j\mathbb E|J_j^*|^2
 +\log^2(eN_{\max})\sum_j\mathbb E(y_j^*)^2+Y(c+2)^2\right)\\
&\le O\!\left((2X/Y+1)X+Y\log^2(eN_{\max})+Y(c+2)^2\right)\\
&\le O\!\left(X^2/Y+X+Y\log^2(e/\rho)+C^2Y\right).
\end{aligned}
\label{eq:hint-profile-moments}
\end{equation}

Define a random variable $T$ as $\forall 1\le s\le \ell,\Pr[T=s-1]=Q_s/\sum_tQ_t$. Lemma~\ref{lem:entropy-mean} gives
\[
\begin{aligned}
\mathsf H(T)
&\le\log\!\left(e\left(1+\frac{\sum_s(s-1)Q_s}{\sum_sQ_s}\right)\right)\\
&\le\log\!\left(e\left[1+O\!\left(X^2/Y^2+X/Y+\log^2(e/\rho)+C^2\right)\right]\right)\\
&\le\log\!\left(e\left[1+O(\rho^{-2}+C^2)\right]\right)\\
&=O\bigl(\log(C+1)+\log(e/\rho)\bigr),
\end{aligned}
\]
where the second step used \eqref{eq:hint-profile-moments} and $\sum_sQ_s\ge Y$ in \eqref{eq:lemma-5-8-bound-of-Qs}, then the third step used $X/Y\le N/Y=1/\rho$.

We use $2^{-c}\le2^{-C}\rho^2$ and \eqref{eq:lemma-5-8-bound-of-Qs}, obtaining
\begin{equation}
\label{eq:hint-outside-mass}
\begin{aligned}
\sum_s2^{-c}Q_s
&\le2^{-C}\rho^2\bigl[X+O(C h_N(Y))\bigr]\\
&\le2^{-C}\bigl[\rho Y+O(C\rho^2Y\log(e/\rho))\bigr]\\
&\le2^{-C}O(C\rho Y)\le\rho Y/8,
\end{aligned}
\end{equation}
where $X\le Y/\rho$, $h_N(Y)=Y\log(e/\rho)$,
$\rho\log(e/\rho)=O(1)$, and $C$ is sufficiently large.
As $\log\frac{eY}{2^{-c}\sum_sQ_s}
\le \log\frac{eY}{2^{-c}Y}\le c+O(1)\le C+2\log(1/\rho)+O(1)$, the first claim of Lemma \ref{lem:entropy-mass} now gives
\[
\begin{aligned}
\sum_{s}h_Y(2^{-c}Q_s)
&=\left(2^{-c}\sum_sQ_s\right)
 \left[\log\frac{eY}{2^{-c}\sum_sQ_s}+\mathsf H(T)\right]\\
&\le2^{-C}\rho^2\bigl[X+O(C h_N(Y))\bigr]\cdot
 O(C+\log(C+1)+\log(e/\rho))\\
&\le2^{-C}C^2O\!\left(\rho^2[X+h_N(Y)]\log(e/\rho)\right).
\end{aligned}
\]
Using $X\le Y/\rho$ and $\rho\log(e/\rho)=O(1)$\footnote{The function $f(\rho)=\rho\log(e/\rho)$ is increasing on $(0,1]$, since $f'(\rho)=-\log\rho\ge0$, so $f(\rho)\le f(1)=\log e$.}, we have
\[
\rho^2X\log(e/\rho)
\le\rho Y\log(e/\rho)=\rho h_N(Y)\le h_N(Y),\quad
\rho^2h_N(Y)\log(e/\rho)=O(h_N(Y)).
\]
Hence, for sufficiently large $C$,
\[
\sum_sh_Y(2^{-c}Q_s)
\le2^{-C}O(C^2h_N(Y))\le h_N(Y)/8.
\]
\end{proof}

The next lemma gives the space bound needed for induction.
\begin{lemma}[Child space and density contraction]
\label{lem:hint-mean-contraction}
\label{lem:hint-local-cost}
The child capacities of a recursive node satisfy
\begin{equation}
\sum_sX_s\le X+O(C h_N(Y)),\qquad
\sum_sh_Y(Y_s)\le\tfrac14h_N(Y),\qquad
\sum_s \rho_s\le \rho/4.
\label{eq:hint-rounded-sums}
\end{equation}
Consequently,
\begin{equation}
\sum_{s\ \mathrm{recursive}}\bigl(X_s+C^2 h_Y(Y_s)\bigr)
+\sum_{s\ \mathrm{direct}}X_s
+\sum_{s\ \mathrm{cheap}}\operatorname{Len}_{\rm CR}(s)
\le X+C^2 h_N(Y),
\label{eq:hint-local-cost}
\end{equation}
where $\operatorname{Len}_{\rm CR}(s)$ is the CR sketch length at
child $s$. (Small-instance leaves are charged separately in the proof of Theorem~\ref{thm:hint-forest}.)
\end{lemma}
\begin{proof}
Put $b=2\eta(D+1)+1$. By \eqref{eq:hint-public-children}, $X_s\le Q_s+b,Y_s\le2^{-c}Q_s+b$.

We first bound the total contribution of these margins.
Since $m=Y$, we have $X\le Y\log^3n$ in a recursive node and hence $b\le3\eta(X+Y+1)=O(\eta Y\log^3n)$.

As $\ell=O(A\log^{14}n),\eta=\log^{-A}n$ and $\log(e/\eta)=O(A\log n)$,
\[
\ell b\log(e/\eta)
\le O\bigl(A^2Y\log^{18-A}n\bigr)
\le\frac{Y}{16\log^{13}n}\le\frac{\rho Y}{8}
\]
for sufficiently large $A$.\footnote{\label{ftnt:111}As the first stopping rule does not apply, there is $N<D\log^{10}n\le 2Y\log^{13}n$. Thus $\rho\ge 1/(2\log^{13}n)$.}
Since $\rho\le 1\le \log(e/\eta)$ and $b\ge\eta Y$, this controls both margin terms:
\[
\ell b\le\rho Y/8,\qquad
\ell h_Y(b)=\ell b\log(eY/b)
\le\ell b\log(e/\eta)\le\rho Y/8\le Y\log(e/\rho)/8= h_N(Y)/8.
\]

Also, by \eqref{eq:hint-outside-mass},
\[
2^{-c}Q_s+b\le\sum_t2^{-c}Q_t+\ell b\le\rho Y/4\le Y.
\]
Now \eqref{eq:lemma-5-8-bound-of-Qs},
Lemma~\ref{lem:hint-entropy-cost}, Lemma~\ref{lem:entropy-mass} and $\rho Y\le Y\le h_N(Y)$ give
\[
\begin{aligned}
\sum_sX_s
&\le\sum_sQ_s+\ell b
\le X+O(C h_N(Y))+\rho Y/8
= X+O(C h_N(Y)),\\
\sum_sh_Y(Y_s)
&\le\sum_sh_Y(2^{-c}Q_s+b)\le\sum_sh_Y(2^{-c}Q_s)+\ell h_Y(b)\\
&\le h_N(Y)/8+h_N(Y)/8=\tfrac14h_N(Y).
\end{aligned}
\]
Consequently,
\[
\sum_s\rho_s=\frac{\sum_sY_s}{Y}
\le\frac{\sum_s2^{-c}Q_s+\ell b}{Y}
\le\frac{\rho Y/8+\rho Y/8}{Y}=\rho/4.
\]
This proves \eqref{eq:hint-rounded-sums}.

In each cheap leaf $s$, there must be $D_s\le Y/\log^{10}n$, hence CR sketch costing $O(Y/\log^8n)$ bits.
There are $\ell=O(C\log^3n)$ children, thus their total cost is $O(CY/\log^5n)\le O(CY)\le O(Ch_N(Y))$.
The left side of \eqref{eq:hint-local-cost} is at most
\[
\begin{aligned}
\sum_sX_s+C^2\sum_sh_Y(Y_s)+O(Ch_N(Y))
&\le X+C^2h_N(Y)/4+O(Ch_N(Y))\\
&\le X+C^2h_N(Y)
\end{aligned}
\]
for sufficiently large $C$.
\end{proof}

\subsection{Closing the induction}
\label{sec:hint-forest-cost}

\begin{lemma}[Size and depth of the recursion]
\label{lem:hint-tree-size}
The {\upshape\HintSketch} recursion has $O(\log\log n)$ levels and
$O(\log^{13}n)$ recursive nodes. Including their children, there are
$O(A\log^{27}n)$ nodes.
\end{lemma}
\begin{proof}
By Lemma~\ref{lem:hint-local-cost}, at depth $h$, the total density
is at most $(1/4)^h$. Every recursive node has density at least
$1/(2\log^{13}n)$ (see footnote \ref{ftnt:111}), so there are at most $2\log^{13}n\,(1/4)^h$
such nodes at that depth (Figure~\ref{fig:hint-density}).

\input{2609-version/fig-hint-density}

Summing over depths gives $O(\log^{13}n)$ recursive nodes.
Additionally, the bound drops below one after $O(\log\log n)$ levels.
As $\ell=O(A\log^{14}n)$, including their children gives $O(A\log^{27}n)$ nodes.
\end{proof}

\begin{proof}[Proof of Theorem~\ref{thm:hint-forest}]
Take $\alpha=C^2$, with $A,C$ fixed as above.

\emph{Length.}
If the root is a cheap leaf, its sketch costs $D\log^2N \le O(N/\log^8n)$ bits.
Otherwise, sum the costs over the recursion tree. Write
$\operatorname{Len}_{\rm direct}(v)$ for the length at a direct
leaf $v$, including its residual CR sketch, and
$\operatorname{Len}_{\rm CR}(v)$ for the CR sketch length at any other node.
Then
\[
\begin{aligned}
\operatorname{Len}
={}\sum_{v\ \mathrm{cheap}}\operatorname{Len}_{\rm CR}(v)
 +\sum_{v\ \mathrm{direct}}\operatorname{Len}_{\rm direct}(v)
 +\sum_{v\ \mathrm{small}}\operatorname{Len}_{\rm CR}(v)
 +\sum_{v\ \mathrm{recursive}}\operatorname{Len}_{\rm CR}(v).
\end{aligned}
\]

We use $(N_v,X_v,Y_v)$ and $D_v:=\min(N_v,X_v+Y_v)$ to denote the parameters at node $v$.

\textbf{1)} At a direct leaf, $D_v\le2X_v$ and the small-instance stopping rule
has failed, so $C,\lambda_*\ge 1\ge \eta$ gives
\[
X_v\ge D_v/2>
\frac{C^3\lambda_*\log^6n}{2\eta^2}\ge\log^6n.
\]
Thus \eqref{eq:hint-direct-storage}, including the residual sketch, gives
\[
\begin{aligned}
\operatorname{Len}_{\rm direct}(v)
&\le X_v+O(CX_v/\log n)+O(C\log^3n)
      +O((\eta X_v+1)\log^2n)\\
&\le (1+O(C/\log n))X_v.
\end{aligned}
\]
\textbf{2)} At a recursive node, the same stopping condition gives $\eta D_v\ge1$,
so $t_v=O(\eta D_v)$ and
$\operatorname{Len}_{\rm CR}(v)=O(\eta D_v\log^2n)$.
\textbf{3)}
At a small-instance leaf, $D_v\le C^3\lambda_*\log^6n/\eta^2$,
so its CR sketch costs $O(C^3\lambda_*\log^{2A+8}n)$ bits.
Lemma~\ref{lem:hint-tree-size} bounds the number of such leaves by
$O(A\log^{27}n)$. Hence
\[
\begin{aligned}
\sum_{v\ \mathrm{small}}\operatorname{Len}_{\rm CR}(v)
&\le O(A\log^{27}n)\cdot O(C^3\lambda_*\log^{2A+8}n)\\
&\le O(AC^4\lambda\log^{2A+36}n)
 =\lambda\log^{O_{A,C}(1)}n,
\end{aligned}
\]
where we used $\lambda_*=\lambda+C\ceil{\log\log(n+2)}+C=O(C\lambda\log n)$.

Lemma~\ref{lem:hint-local-cost} bounds the children's charges by
$X_v+C^2h_{N_v}(Y_v)$.
Applying this bound from the root down gives
\[
\sum_{v\ \mathrm{cheap}}\operatorname{Len}_{\rm CR}(v)
+\sum_{v\ \mathrm{direct}}X_v
\le X+C^2h_N(Y).
\]
At each depth, $\sum_{v\ \mathrm{recursive}}(X_v+C^2h_{N_v}(Y_v))$ can also be bounded by $X+C^2h_N(Y)$.
Since $D_v\le X_v+C^2h_{N_v}(Y_v)$, Lemma~\ref{lem:hint-tree-size} therefore gives
\[
\sum_{v\ \mathrm{recursive}}D_v
\le O(\log\log n)\bigl(X+C^2h_N(Y)\bigr).
\]
Substituting these bounds into the total length yields
\[
\begin{aligned}
\operatorname{Len}
&\le (1+O(C/\log n))
 \left(\sum_{v\ \mathrm{cheap}}\operatorname{Len}_{\rm CR}(v)
       +\sum_{v\ \mathrm{direct}}X_v\right)\\
&\quad+O(\eta\log^2n)\sum_{v\ \mathrm{recursive}}D_v
       +\lambda\log^{O_{A,C}(1)}n\\
&\le \bigl(1+O(C/\log n)+O(\eta\log^2n\log\log n)\bigr)
       \bigl(X+C^2h_N(Y)\bigr)
       +\lambda\log^{O_{A,C}(1)}n\\
&\le (1+O(1/\log n))\bigl(X+\alpha h_N(Y)\bigr)
       +\lambda\log^{O(1)}n.
\end{aligned}
\]
Together with the cheap-root case\footnote{In fact, the $O(N/\log^{8}n)$ term occurs only in this case.}, this proves
\eqref{eq:hint-forest-length}.

\emph{Recovery.}
Theorem~\ref{thm:decoder-composition} guarantees exact recovery
whenever all capacity bounds hold. Each bound fails with probability
at most $2^{-\lambda_*}$, and there are $O(A\log^{27}n)$ such bounds.
A union bound therefore gives
\[
\Pr[\text{decoding fails}]
\le O(A\log^{27}n)\,2^{-\lambda_*}
\le 2^{-\lambda},
\]
by the choice of $C$ in $\lambda_*$.
Obviously, all stored maps are binary linear.

\emph{Computation.}
An update changes one row and at most one coordinate per child.
The $\log^{O_A(1)}n$ nodes and columns give $\log^{O_A(1)}n$ column locality
and update time.

For decoding, a row tries at most $(N_{\max}+1)^{B_Y}=\exp(O(C(\log\log n)^2))$ possible
supports outside the hint. For each choice, it solves a binary linear
system for the hinted bits. Since each row and its measurements have
size $\log^{O(1)}n$, processing all prefixes takes
$\exp(O(C(\log\log n)^2))$ time per row.
There are at most $N$ rows per node and $\log^{O(1)}n$ nodes, giving total time
\[
N\log^{O(1)}n\cdot\exp(O(C(\log\log n)^2))
=N\exp(O((\log\log n)^2)).
\]
Forming syndromes and decoding the CR sketches take only
$N\log^{O(1)}n$ additional time.

Finally, we stop any CR call that exceeds the time bound in Proposition~\ref{prop:cr-sketch} and report failure, so the running-time bound also holds when recovery fails.
\end{proof}

%% file: 2609-version/fig-hint-budget.tex
\begingroup
\definecolor{budgetLine}{RGB}{83,101,117}
\definecolor{budgetHint}{RGB}{238,238,238}
\definecolor{budgetNew}{RGB}{237,233,222}
\definecolor{budgetOutside}{RGB}{188,208,226}
\edrev
\begin{figure}[!htbp]
\centering
\begin{tikzpicture}[x=1cm,y=1cm,font=\figurefontsize,text=black,
arr/.style={-{Latex[length=2mm]},draw=budgetLine,semithick}]
\node[anchor=east] at (0,.35) {parent};
\filldraw[draw=budgetLine,fill=budgetHint] (.25,0) rectangle (4.15,.7);
\filldraw[draw=budgetLine,fill=budgetOutside] (4.15,0) rectangle (13.5,.7);
\node at (2.2,.35) {hint budget $X$};
\node at (8.8,.35) {outside budget $C^2 h_N(Y)$};
\node[anchor=east] at (0,-1.3) {children};
\filldraw[draw=budgetLine,fill=budgetHint] (.25,-1.85) rectangle (4.15,-.75);
\filldraw[draw=budgetLine,fill=budgetNew] (4.15,-1.85) rectangle (7.6,-.75);
\filldraw[draw=budgetLine,fill=budgetOutside] (7.6,-1.85) rectangle (10.9,-.75);
\draw[draw=budgetLine,dashed] (10.9,-1.85) rectangle (13.5,-.75);
\node at (2.2,-1.3) {inherited cost $X$};
\node[font=\figurefontsize,align=center] at (5.875,-1.3) {new hints\\$O(C h_N(Y))$};
\node at (9.25,-1.3) {$\le C^2 h_N(Y)/4$};
\node[font=\figurefontsize,align=center] at (12.2,-1.3) {remaining\\budget};
\draw[arr] (2.2,0)--(2.2,-.75);
\draw[arr] (8.8,0)--(8.8,-.75);
\draw[arr] (6,0)--(5.875,-.75);
\node[anchor=north,font=\figurefontsize] at (6.8,-2.15)
  {Choose $C$ so that the released outside budget pays for the new hints.};
\end{tikzpicture}
\makeatletter
\let\ai@originalcaption\@makecaption
\long\def\@makecaption#1#2{{\edrev\ai@originalcaption{#1}{#2}}}
\makeatother
\caption{\edrev Why hints are paid for only once. The same $X$ passes
to the children; the shrinking outside cost funds the added hints.
The schematic omits rounding and small repair costs.}
\label{fig:hint-budget}

\end{figure}
\edrev
\endgroup

%% file: 2609-version/fig-hint-density.tex
\begingroup
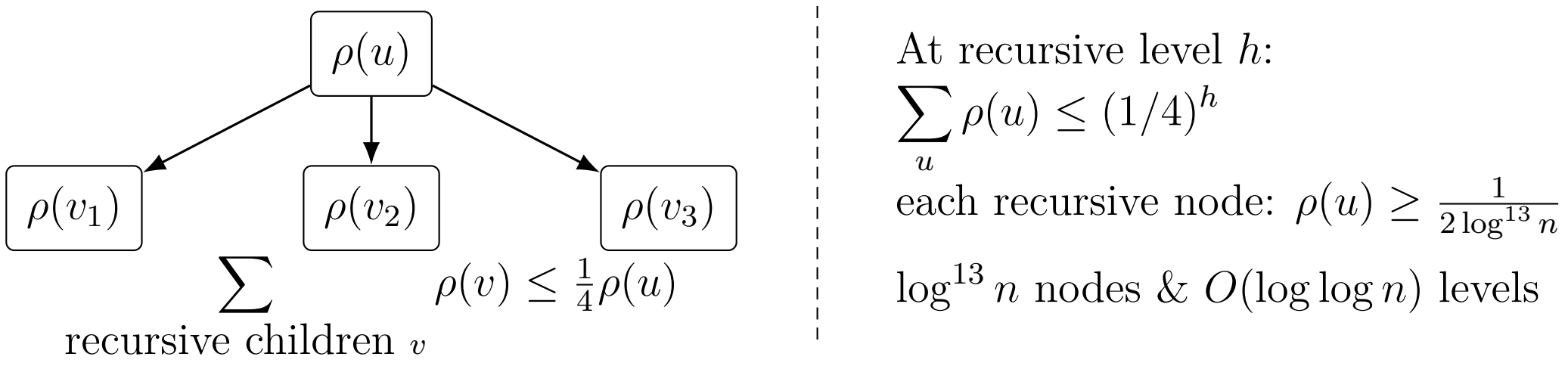
\begin{figure}[H]
\centering
\begin{tikzpicture}[x=1cm,y=1cm,font=\figurefontsize,text=black,
box/.style={draw=black,rounded corners=2pt,inner sep=5pt},
arr/.style={-{Latex[length=2mm]},draw=black,semithick}]
\node[box] (p) at (3.1,.7) {$\rho(u)$};
\node[box] (a) at (.5,-.65) {$\rho(v_1)$};
\node[box] (b) at (3.1,-.65) {$\rho(v_2)$};
\node[box] (c) at (5.7,-.65) {$\rho(v_3)$};
\draw[arr] (p)--(a);\draw[arr] (p)--(b);\draw[arr] (p)--(c);
\node[align=center] at (3.1,-1.5)
  {$\displaystyle\sum_{\text{\figurefontsize recursive children }v}\rho(v)\le\tfrac14\rho(u)$};
\draw[black,densely dashed] (7,-1.8)--(7,1.15);
\node[anchor=west] at (7.55,.75) {At recursive level $h$:};
\node[anchor=west] at (7.55,.05)
  {$\displaystyle\sum_u\rho(u)\le(1/4)^h$};
\node[anchor=west] at (7.55,-.65)
  {each recursive node: $\rho(u)\ge\frac1{2\log^{13}n}$};
\node[anchor=west] at (7.55,-1.35)
  {$\log^{13}n$ nodes \& $O(\log\log n)$ levels};
\end{tikzpicture}
\makeatletter
\let\ai@originalcaption\@makecaption
\long\def\@makecaption#1#2{{\ai@originalcaption{#1}{#2}}}
\makeatother
\caption{ Density controls the width as well as the depth of
the recursion, where $\rho(u)=Y(u)/N(u)$. Each recursive node consumes
a minimum share of a total density that decreases geometrically.}
\label{fig:hint-density}

\end{figure}
\endgroup

%% file: 2609-version/6toplevel.tex
\section{Top-level analysis and the main theorem}
\label{sec:analysis-top}

\subsection{Proof overview}
\label{sec:top-proof-overview}

In this section, we use the HintSketch guarantee in Theorem~\ref{thm:hint-forest} as a black box and build the top layer analysis.
It remains to bound the budgets of the top layer's children and the cost of its final CR sketch.

Note that the main term in Theorem~\ref{thm:hint-forest} is $X+\alpha h_N(Y)$.
Let $w$ be the number of bits in a Pinsketch word, and index a child
by its word $a$ and bit $k$.
As in Definition~\ref{def:exact-child}, let $U_{a,k}$ and $V_{a,k}$ count workers returning $\perp$ and predicting incorrectly, respectively.
Ignoring concentration margins for now, we therefore want
\begin{equation}
\sum_{a,k}X_{a,k}\le dw,\qquad\sum_{a,k}\alpha h_d(Y_{a,k})=O(d).
\end{equation}

We choose two families of bounds, $q_a$ and $p_k$, and prove
\begin{equation}
\label{eq:sec6-proof-overview-bound}
\mathbb EU_{a,k}\le q_a,\qquad \mathbb EV_{a,k}\le q_ap_k,\qquad
\sum_aq_a= d,\qquad \sum_kp_k=1.
\end{equation}
Taking $X_{a,k}=q_a$ and $Y_{a,k}=q_ap_k$ then gives
$\sum_{a,k}X_{a,k}=w\sum_aq_a\le dw$.
For the second bound, we use Lemma~\ref{lem:entropy-mass} to obtain $\sum_{a,k}\alpha h_d(q_ap_k)=O(d)$. Thus the children's main length terms
sum to $dw+O(d)$, while the concentration margins add only $O(d/\log n)$.

\subsection{Errors across words and within a word}
\label{sec:top-child-parameters}
\label{sec:top-row-analysis}

We use the fixed constant $A$ from Section~\ref{sec:hint-row-parameters}
and the fixed coefficient $\alpha$ from Theorem~\ref{thm:hint-forest}.
So in this section we simply use $A,\alpha=O(1)$.
Choose a separate sufficiently large constant $C'\ge A$ and set
\[
\eta=\log^{-A}n,\qquad
\lambda_*=\lambda+C'\ceil{\log\log(n+2)}+C'.
\]
For $d\ge n/2$, we just directly store $x$.
Otherwise, for $d\le (C')^3\lambda_*\log^6n/\eta^2$, use
$\CR_{n,d}$. In the remaining regime use $m=d$ rows and
\[
N=\ceil{n/d},\qquad w=\ceil{\log(N+1)},\qquad
B=\min\{N,\ceil{C'\log(\log n+2)}\},\qquad \ell=Bw.
\]
For each child, apply
Theorem~\ref{thm:hint-forest} with $\lambda$ replaced by
$\lambda+\ceil{\log(2\ell)}$, making its failure probability at most
$2^{-\lambda-1}/\ell$.

We use $(a,k)$ to denote column $s=(a-1)w+k$, which is $k$-th bit of word $a$.
We now begin to bound the expectations of $U$ and $V$.

\begin{lemma}[Across words]
\label{lem:top-word-bounds}
Place $d$ reference labels uniformly at distinct positions in $[n]$,
let $T_j$ count the labels in row $j$, and set
$q_a=\sum_j\Pr[T_j\ge a]$ for $a\ge1$. Then
\[
\mathbb E U_{a,k}\le q_a,\qquad
\sum_{a\ge1}q_a=d,\qquad \mathsf H((q_a/d)_{a\ge 1})\le\log(2e).
\]
\end{lemma}
\begin{proof}
As a row with $\le a-1$ nonzeros can be recovered from the first $a-1$ words by Pinsketch, only rows with at least $a$ nonzeros can contribute to $U_{a,k}$ or $V_{a,k}$.
To bound the number of such rows, extend the actual support of $x$ to exactly $d$ positions and apply the same random permutation.
The enlarged set is uniformly distributed among all $d$-subsets of $[n]$, giving that
\[
\mathbb E U_{a,k}
\le \sum_j\Pr[\text{row }j\text{ has at least }a\text{ nonzeros}]
\le \sum_j\Pr[T_j\ge a]
=q_a.
\]
By definition, $T_j$ is hypergeometric, with means $\mathbb ET_j=dN_j/n\le2$, so
\[
\begin{aligned}
\sum_{a\ge1}q_a&=\sum_j\mathbb ET_j=d,\\
\sum_{a\ge1}(a-1)q_a
&=\frac12\sum_j\mathbb E[T_j(T_j-1)]
\le\frac12\sum_j(\mathbb ET_j)^2\le d.
\end{aligned}
\]
The second equation uses $\mathbb E[T_j(T_j-1)]\le (\mathbb E T_j)^2+\mathrm{Var}(T_j)-\mathbb E T_j\le (\mathbb E T_j)^2$ as $\mathrm{Var}(T_j)\le \mathbb E[T_j]$ for hypergeometric variables. Thus, the distribution $q_a/d$ on $a-1$ has mean at most one. Lemma~\ref{lem:entropy-mean} hence gives its entropy bound.
\end{proof}

\begin{lemma}[Within a word]
\label{lem:top-bit-bounds}
Conditional on a row's word prediction being incorrect, its first
incorrect bit has distribution
\begin{equation}
p_k=\frac{2^{w-k}}{2^w-1},\quad 1\le k\le w,
\qquad \sum_{k=1}^w p_k=1,\qquad \mathsf H((p_k)_{k=1}^w)\le2.
\label{eq:top-first-error}
\end{equation}
For every column $(a,k)$, $\mathbb E V_{a,k}\le q_ap_k$.
\end{lemma}
\begin{proof}
By Algorithm~\ref{alg:construction-top-row}, worker $j$ uses the
first $a-1$ recovered words to choose a candidate support, then uses
that candidate to predict word $a$.
Both the predicted and true field elements are multiplied by an independent random scalar $\alpha_{j,a}$.

If its unscaled prediction is incorrect, the difference from the true value is
nonzero, so multiplication by $\alpha_{j,a}$ makes the difference
uniform over all nonzero $w$-bit vectors.
The first nonzero bit is at position $k$ when the first $k-1$ bits
are zero and bit $k$ is one, proving that the distribution is
\[
p_k=\frac{2^{w-k}}{2^w-1},
\]
which is a truncated geometric distribution with entropy at most two.

The algorithm returns $\perp$ for the rest of the word after its
first wrong bit, so each row makes at most one incorrect prediction
per word. By Lemma~\ref{lem:top-word-bounds}, only rows with
$T_j\ge a$ can make such an error. Therefore,
\[
\mathbb E V_{a,k}
\le p_k\sum_j\Pr[T_j\ge a]
=q_ap_k.
\]
\end{proof}

\input{2609-version/fig-top-budget}

As $q_a,p_k$ can be computed from the construction parameters independently of $x,J$, we can add the same margin to both expectation bounds and choose
\begin{equation}
X_{a,k}=\min\{d,\ceil{q_a+\eta(d+1)}\},\qquad
Y_{a,k}=\min\{d,\ceil{q_ap_k+\eta(d+1)}\}.
\label{eq:top-public-children}
\end{equation}

\begin{lemma}[Top-layer child capacities]
\label{lem:top-initial-budget}
For the capacities in \eqref{eq:top-public-children}, each of the events
$U_{a,k}>X_{a,k}$ and $V_{a,k}>Y_{a,k}$ has probability at most
$2^{-\lambda_*}$. Moreover,
\begin{equation}
\begin{aligned}
\sum_{a,k}X_{a,k}&\le dw+O(C'd/\log n),\\
\sum_{a,k}h_d(Y_{a,k})&\le O(d)+O(C'd/\log n).
\end{aligned}
\label{eq:top-initial-budget}
\end{equation}
Consequently,
\[
\sum_{a,k}\bigl(X_{a,k}+\alpha h_d(Y_{a,k})\bigr)
\le dw+O(d)+O(C'd/\log n).
\]
\end{lemma}
\begin{proof}
We first check that the capacities are sufficient.
Lemma \ref{lem:top-word-bounds} and Lemma \ref{lem:top-bit-bounds} give $\mathbb E U_{a,k}\le q_a$ and
$\mathbb E V_{a,k}\le q_ap_k$, while we add $\eta(d+1)$ to these bounds in \eqref{eq:top-public-children}.
Lemma~\ref{lem:public-capacities}, with $D=m=d$ and
$\Delta=\eta(d+1)$, therefore bounds each violation probability by
\[
\exp\!\left(-\frac{2\eta^2(d+1)^2}{5d}\right)
\le 2^{-\lambda_*},
\]
since $d>(C')^3\lambda_*\log^6n/\eta^2$.

Next we bound the space. Write $b=\eta(d+1)+1=O(\eta d)$ for
the maximum amount added to each capacity. By
Lemma~\ref{lem:entropy-mass},
\begin{equation}
\begin{aligned}
\sum_{a,k}X_{a,k}
&\le w\sum_{a=1}^B q_a+\ell b
\le dw+\ell b,\\
\sum_{a,k}h_d(Y_{a,k})
&\le \sum_{a=1}^B\sum_{k=1}^w h_d(q_ap_k)+\ell h_d(b)\\
&\le \sum_{a,k} q_ap_k(\log e+\log(d/q_a)+\log(1/p_k))+\ell h_d(b)\\
&\le d\bigl[\log e+\mathsf H(q/d)+\mathsf H(p)\bigr]
     +\ell h_d(b)\\
&=O(d)+\ell h_d(b).
\end{aligned}
\label{eq:top-ideal-entropy}
\end{equation}
Here the first inequality for $h_d$ uses its monotonicity and
subadditivity, while the last two lines use Lemma \ref{lem:top-word-bounds} and Lemma \ref{lem:top-bit-bounds}. Since
$\ell=O(C'\log^2n)$ and $\eta=\log^{-A}n$,
\[
\ell h_d(b)
=\ell b\log(ed/b)
\le O\!\left(C'A\eta d\log^3n\right)= O\!\left(C'dA\log^{3-A}n\right)
=O(C'd/\log n).
\]
W.l.o.g. we can assume $A\ge 10$, which gives the last step.
Substituting this into \eqref{eq:top-ideal-entropy} proves
\eqref{eq:top-initial-budget}.
\end{proof}

It remains to cover the residual weight $Z$ with the final CR sketch.
Only rows containing more than $B$ nonzeros can leave errors.

\begin{lemma}[Residual repair]
\label{lem:top-residual-repair}
For the capacity
\begin{equation}
t=\min\{n,d(B+1),\ceil{2\eta(d+1)}\},
\label{eq:top-residual-capacity}
\end{equation}
we have $\Pr[Z>t]\le2^{-\lambda_*}$, and the final CR sketch uses
$O(d/\log n)$ bits.
\end{lemma}
\begin{proof}
When $B=N$, there must be $\Pr[Z>t]=0$. So we assume that $B<N$.
In our algorithm, rows with at most $B$ nonzeros can be recovered exactly.
In a row with $r>B$ nonzeros, the decoder outputs at most $B$
positions, so the residual weight is at most $r+B\le2r$.
Thus $Z$ is at most twice the number of nonzeros lying in rows with more than $B$ nonzeros.

Fix a coordinate $i$ such that $x_i=1$ and consider its row. The number of other nonzeros $H$ in that row is hypergeometric with mean at most two, as $m=d$. By Lemma~\ref{lem:balanced-load} and the choice of $B$, the row has more than $B$ nonzeros with probability 
$$
\mathrm{Pr}[1+H>B]=\Pr[H\ge B]\le e^{-B/8}\le (\log n+2)^{-C'/(8\ln 2)}\le \log^{-2A}n=\eta^2,
$$
where we use $C'$ sufficiently large such that $C'\ge 16A$. Summing over the at most $d$ support coordinates gives $\mathbb EZ\le2d\eta^2\le\eta(d+1)$.
Lemma~\ref{lem:public-capacities}, with $D=m=d$, $K=B$, and
$\Delta=\eta(d+1)$, therefore gives
\[
\Pr[Z>2\eta(d+1)]\le\exp\left(-\dfrac{\eta^2d}{10B^2}\right)\le \exp\left(-\Omega(C'\lambda_*\log^4 n)\right)\le 2^{-\lambda_*},
\]
where $d\ge (C')^3\lambda_*\log^6 n/\eta^2$ and $B=O(C'\log n)$ is used.
Since also $Z\le\min\{n,2d\}$, the capacity $t$ satisfies
$\Pr[Z>t]\le2^{-\lambda_*}$.
Finally, the final CR sketch uses $O(t\log^2n)=O(\eta d\log^2n)=O(d/\log n)$ bits.
\end{proof}

\subsection{The main theorem}
\label{sec:complete-main}
\label{sec:top-residual}
\label{sec:complete-resources}

This completes the construction and gives the following guarantee.

\begin{theorem}[Main guarantee]
\label{thm:complete-main}
For $2\le d\le n$ and $0<\eps\le1/2$, let $\lambda = \ceil{\log(1/\eps)}$. There is a public-coin
binary linear sketch recovering each fixed $x\in\mathcal S_{n,d}$
with probability at least $1-\eps$, using at most
\begin{equation}
L(n,d)+O(d)+\log^{O(1)} n\,\lambda
\label{eq:complete-main-length}
\end{equation}
bits. Column locality and update time are $\log^{O(1)} n$.
Worst-case decoding takes $d\exp(O((\log\log n)^2))$ time,
which is $d\,n^{o(1)}$.
\end{theorem}
\begin{proof}
We now fix $C'$ sufficiently large for the preceding lemmas.

\emph{Recovery.}
In our parameter choice, each valid child call fails with probability at most $2^{-\lambda-1}/\ell$. Combining this with
Lemmas~\ref{lem:top-initial-budget} and~\ref{lem:top-residual-repair}
and Theorem~\ref{thm:decoder-composition}, a union bound gives
\[
\Pr[\text{recovery fails}]
\le(2\ell+1)2^{-\lambda_*}
 +\ell\frac{2^{-\lambda-1}}{\ell}
\le2^{-\lambda}\le\eps.
\]
The cases of $d\ge n/2$ and $d\le (C')^3\lambda_*\log^6n/\eta^2$ are deterministic and error-free.
Obviously, all stored maps are binary linear.

\emph{Length.}
Write $\operatorname{Len}_s$ for the length of child $s$ and
$\operatorname{Len}_{\rm CR}$ for that of the final CR sketch.
Theorem~\ref{thm:hint-forest} and
Lemmas~\ref{lem:top-initial-budget} and~\ref{lem:top-residual-repair} give
\[
\begin{aligned}
\operatorname{Len}
&=\operatorname{Len}_{\rm CR}+\sum_{s=1}^{\ell}\operatorname{Len}_s\\
&\le (1+O(1/\log n))\sum_{s=1}^{\ell}\bigl(X_s+\alpha h_d(Y_s)\bigr)\\
&\quad+O(d/\log n+\ell d/\log^8n)+\ell(\lambda+\ceil{\log(2\ell)})\log^{O(1)}n\\
&\le (1+O(1/\log n))\bigl(dw+O(d)+O(C'd/\log n)\bigr)\\
&\quad+O(d/\log n+C'd/\log^6n)+\lambda\log^{O(1)}n\\
&=dw+O(dw/\log n)+O(d)+\lambda\log^{O(1)}n\\
&=L(n,d)+O(d)+\lambda\log^{O(1)}n.
\end{aligned}
\]
Here $\ell=O(C'\log^2n)$, $w=O(\log n)$, and $dw=L(n,d)+O(d)$.\footnote{Since $N=\ceil{n/d}$ and $w=\ceil{\log(N+1)}$,
we have $dw=d\log(n/d)+O(d)=h_n(d)+O(d)=L(n,d)+O(d)$,
where the last equality follows from the estimate in
Section~\ref{sec:notations}.}
For $d\ge n/2$, direct storage uses $n\le L(n,d)+1$ bits, while the small-instance mode uses
\[
O((C')^3\lambda_*\log^{2A+8}n)
\le O((C')^4\lambda\log^{2A+9}n)
=\lambda\log^{O(1)}n
\]
bits, using $\lambda_*=O(C'\lambda\log n)$.

\emph{Updates and decoding.}
A toggle changes one row and at most one input bit in each of the
$\ell=\log^{O(1)}n$ columns. Theorem~\ref{thm:hint-forest} and
Proposition~\ref{prop:cr-sketch} then give polylogarithmic column locality and update time.

Row decoding, residual-syndrome formation, and CR recovery together
cost $d\log^{O(1)}n$: the row outputs contain at most $dB$
coordinates, and $t\le d(B+1)$.
As the dominant term, the $\ell$ child calls take $d\exp(O((\log\log n)^2))$ time by
Theorem~\ref{thm:hint-forest}.
\end{proof}
\edrev

\begin{corollary}[Near-counting-bound space]
\label{cor:near-counting}
Fix constants $A_0,c_0>0$. If
$n\le2^{(\log d)^{A_0}}$ and $\lambda\le d^{1-c_0}$,
the sketch uses at most $L(n,d)+O_{A_0,c_0}(d)$ bits, with
$\log^{O_{A_0}(1)}d$ column locality and update time,
and $d^{1+o(1)}$ decoding time as $d\to\infty$.
\end{corollary}
\begin{proof}
Substitute $\log n=O((\log d)^{A_0})$:
the additive space is at most
$(d^{1-c_0}+1)\log^{O(A_0)}d=o(d)$, and the decoding
overhead is $\exp(O_{A_0}((\log\log d)^2))=d^{o(1)}$.
\end{proof}

%% file: 2609-version/fig-top-budget.tex
\begingroup
\edrev
\definecolor{zoomInk}{RGB}{0,0,0}
\definecolor{zoomLine}{RGB}{142,151,160}
\definecolor{zoomHint}{RGB}{68,119,170}
\definecolor{zoomAccent}{RGB}{45,85,125}
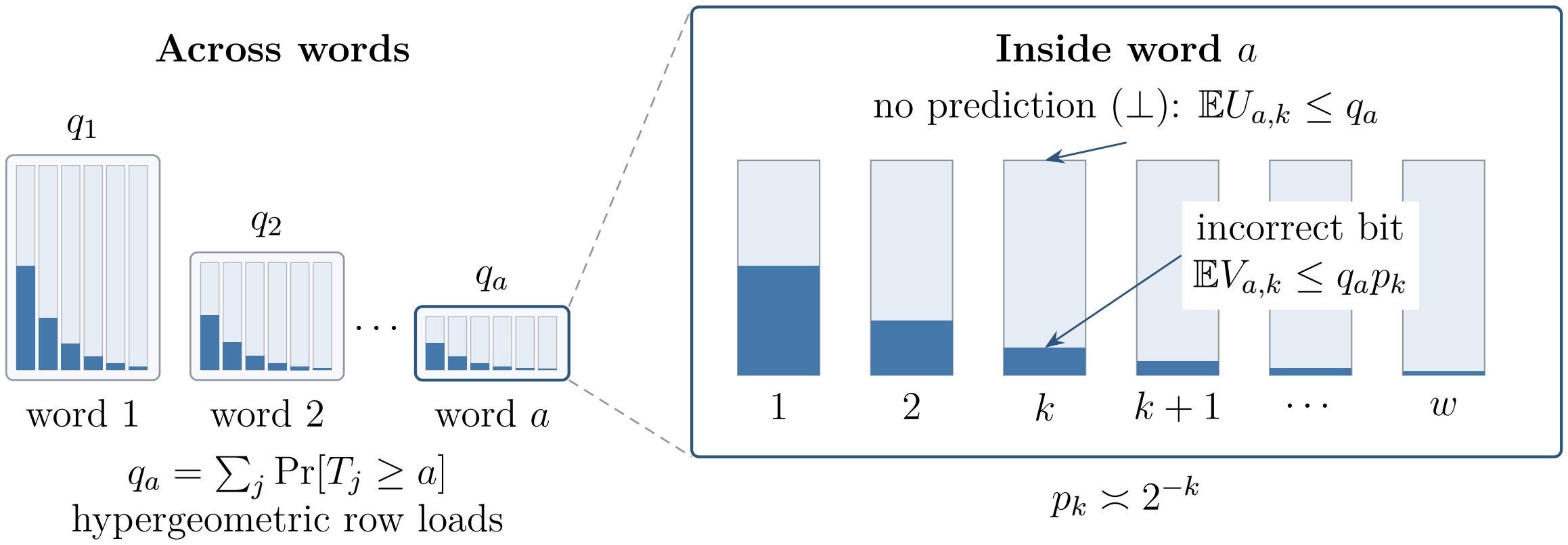
\begin{figure}[!htbp]
\centering
\begin{tikzpicture}[x=1cm,y=1cm,font=\figurefontsize,text=zoomInk,
  >=Stealth]
\node[font=\figurefontsize\bfseries] at (2.7,3.45) {Across words};
\foreach \lab/\x/\h in {1/0/2.0,2/1.8/1.05,a/4.0/.52}{
  \filldraw[fill=zoomHint!5,draw=zoomLine,line width=.5pt,
    rounded corners=2pt] (\x,.2) rectangle ({\x+1.5},{.4+\h});
  \foreach \k in {1,...,6}{
    \filldraw[fill=zoomHint!13,draw=zoomLine!70,line width=.25pt]
      ({\x+.1+.22*(\k-1)},.30)
      rectangle ({\x+.28+.22*(\k-1)},{.30+\h});
    \fill[zoomHint] ({\x+.1+.22*(\k-1)},.30)
      rectangle ({\x+.28+.22*(\k-1)},{.30+(32/63)*\h*pow(.5,\k-1)});
  }
  \node at ({\x+.75},{.67+\h}) {$q_{\lab}$};
  \node[font=\figurefontsize] at ({\x+.75},-.12) {word $\lab$};
}
\node at (3.65,.70) {$\cdots$};
\draw[zoomAccent,line width=.8pt,rounded corners=2pt]
  (4,.2) rectangle (5.5,.92);
\draw[zoomLine,densely dashed,line width=.5pt] (5.5,.92)--(6.7,3.85);
\draw[zoomLine,densely dashed,line width=.5pt] (5.5,.2)--(6.7,-.55);
\node[font=\figurefontsize,align=center] at (2.75,-.94)
  {$q_a=\sum_j\Pr[T_j\ge a]$\\hypergeometric row loads};

\filldraw[fill=white,draw=zoomAccent,line width=.8pt,rounded corners=2pt]
  (6.7,-.55) rectangle (15.2,3.85);
\node[font=\figurefontsize\bfseries] at (10.95,3.45) {Inside word $a$};
\foreach \i/\lab in {0/1,1/2,2/k,3/{k+1},4/{\cdots},5/w}{
  \filldraw[fill=zoomHint!13,draw=zoomLine,line width=.4pt]
    ({7.15+1.30*\i},.25) rectangle ({7.95+1.30*\i},2.35);
  \fill[zoomHint] ({7.15+1.30*\i},.25)
    rectangle ({7.95+1.30*\i},{.25+2.1*(32/63)*pow(.5,\i)});
  \node[font=\figurefontsize] at ({7.55+1.30*\i},-.06) {$\lab$};
}
\node[font=\figurefontsize] (hintlabel) at (10.95,2.86)
  {no prediction ($\perp$): $\mathbb E U_{a,k}\le q_a$};
\draw[->,zoomAccent,line width=.55pt]
  (hintlabel.south)--(10.15,2.35);
\node[font=\figurefontsize,align=center,fill=white,inner sep=3pt] (errorlabel) at (12.65,1.42)
  {incorrect bit\\$\mathbb E V_{a,k}\le q_a p_k$};
\draw[->,zoomAccent,line width=.55pt]
  (errorlabel.west)--(10.15,{.25+2.1*8/63});
\node at (10.95,-.94) {$p_k\asymp 2^{-k}$};
\end{tikzpicture}
\makeatletter
\let\ai@originalcaption\@makecaption
\long\def\@makecaption#1#2{{\edrev\ai@originalcaption{#1}{#2}}}
\makeatother
\caption{\edrev Bounds for the two child capacities. In word $a$, the
light columns show the common bound $q_a$ on the expected number of
rows returning $\perp$; the dark bars show the bounds $q_ap_k$ on
incorrect predictions. The word heights follow the hypergeometric row-load tails,
while $p_k$ decays geometrically within each word.}
\label{fig:top-budget}
\end{figure}
\edrev
\endgroup

%% file: 2609-version/Z-AIdisclosure.tex
\edrev
\section{AI disclosure}
\label{sec:ai-methodology}

The initial algorithmic construction was developed in late 2025 and
refined in early 2026. It achieved $L(n,d)+O(d)$ space, polynomial-time
encoding and decoding, and inverse-polynomial failure probability in
$d$. The authors developed this construction, its key lemmas, and its
proof ideas without substantive AI assistance.

Beginning in mid-2026, we used OpenAI Codex, Google Gemini, and Anthropic Claude to assist
with organizing and checking the proofs. AI assistance also contributed
to incorporating the Cheraghchi--Ribeiro sketch to simplify the proofs,
extending the analysis to general failure parameter $\lambda$, and
refining the analysis to improve the encoding and decoding complexity
bounds. These contributions affect the construction and
analysis in Sections~\ref{sec:construction}--\ref{sec:analysis-top} and
the appendix. We also used AI tools for literature searches, exposition,
and figures; Gemini assisted with drafting the abstract and introduction.
The authors take full responsibility for the mathematical claims and
references. The results do not rely on code, experiments, or computational
verification.
\edrev

%% file: 2609-version/B-probabilistic-tools.tex
\section{Proofs of the probabilistic tools}
\label{app:probabilistic-tools}

We prove the lemmas from Section~\ref{sec:analysis-probability} in the
same order: the probability bounds come first, followed by the entropy
and prefix-tail estimates used to sum child-sketch costs.

\paragraph{ Hypergeometric tails (Lemma~\ref{lem:balanced-load}).}
\begin{proof}
For $0\le r\le q$, counting $r$-subsets of $S\cap I$ gives
$$
\mathbb E\binom Zr=\dfrac{\binom{q}{r}\binom{N-r}{K-r}}{\binom{N}{K}}=\dfrac{1}{r!}\prod_{i=0}^{r-1}\dfrac{(K-i)(q-i)}{N-i}\le \dfrac{\mu^r}{r!},
$$
and in particular
$\mathbb E[Z(Z-1)]\le\mu^2$.
Set $u=e^\theta-1\ge0$ and expand $(1+u)^Z$ to obtain
\begin{equation*}
\mathbb E e^{\theta Z}
=\sum_{r=0}^q u^r\mathbb E\binom Zr
\le\sum_{r\ge0}\frac{(\mu u)^r}{r!}=e^{\mu u}.
\end{equation*}

Markov's inequality with $\theta=\ln(1+\delta)$ gives, for $\delta\ge0$,
\[
\Pr[Z\ge(1+\delta)\mu]
\le\exp\{-\mu((1+\delta)\ln(1+\delta)-\delta)\}
\le\exp\!\left(-\frac{\mu\delta^2}{2(1+\delta)}\right).
\]
For the last step, integrate $\ln(1+s)\ge s/(1+\delta)$ over
$0\le s\le\delta$. Taking $\delta=t/\mu$ proves
\eqref{eq:hypergeometric-tail}.
For $k\ge2\mu$, set $t=k-\mu\ge k/2$ to obtain
$\Pr[Z\ge k]\le\exp(-(k-\mu)^2/(2k))\le e^{-k/8}$.

\end{proof}

\paragraph{ Concentration under a random injection (Lemma~\ref{lem:injection-concentration}).}
\begin{proof}
Reveal $\pi(1),\ldots,\pi(D),R_1,\ldots,R_b$ in this order.
Let $\mathcal F_i$ denote the information revealed after $i$ steps,
and define the Doob martingale $M_i=\mathbb E[F\mid\mathcal F_i]$.
Then $M_0=\mathbb EF$ and $M_{D+b}=F$.

Fix the revealed history before step $i\le D$, and consider two
possible values $y,y'$ of $\pi(i)$. Composing an injection with
the transposition $(y\ y')$ bijects its completions with
$\pi(i)=y$ and those with $\pi(i)=y'$. This preserves the revealed
prefix, since neither position is occupied by an earlier label.
The conditional distributions on these completion sets are uniform.
Couple them through this bijection and use the same $R$ in both.
The assumed bound on $F$ gives
$$
\left|
\mathbb E[F\mid\mathcal F_{i-1},\pi(i)=y]
-\mathbb E[F\mid\mathcal F_{i-1},\pi(i)=y']
\right|\le L_p.
$$
Thus, conditional on $\mathcal F_{i-1}$, the possible values of $M_i$
lie in an interval of length at most $L_p$. The same holds for
$M_i-M_{i-1}$, since $M_{i-1}$ is already determined.

At a subsequent step revealing $R_j$, independence allows us to
compare two values of $R_j$ while keeping all remaining random
variables identical. The corresponding values of $F$ differ by
at most $L_r$. Averaging shows that the conditional range of this
martingale increment has length at most $L_r$.

Every increment has conditional mean zero. Conditional Hoeffding's
lemma therefore gives
$\mathbb E[e^{\theta(M_i-M_{i-1})}\mid\mathcal F_{i-1}]
\le e^{\theta^2c_i^2/8}$,
where $c_i=L_p$ for $i\le D$ and $c_i=L_r$ otherwise.
Iterating conditional expectations yields
$$
\mathbb E e^{\theta(F-\mathbb EF)}
\le \exp\left(\frac{\theta^2}{8}(DL_p^2+bL_r^2)\right).
$$
By Markov's inequality, for every $\theta>0$,
$$
\Pr[F-\mathbb EF\ge t]
\le \exp\left(-\theta t+
              \frac{\theta^2}{8}(DL_p^2+bL_r^2)\right).
$$
Taking $\theta=4t/(DL_p^2+bL_r^2)$ proves
\eqref{eq:injection-concentration}.
\end{proof}

\paragraph{ Separation by random measurements (Lemma~\ref{lem:random-separation}).}
\begin{proof}
Fix $C\ne S$ and choose a coordinate $v\in C\triangle S$.
Since common coordinates cancel over $\F_2$, 
$$
B(C)+B(S)
=\beta(v)+\sum_{u\in(C\triangle S)\setminus\{v\}}\beta(u).
$$
Condition on all vectors other than $\beta(v)$.
The remaining sum is then fixed, while $\beta(v)$ is still uniform
in $\F_2^q$ and independent of those vectors. Exactly one of its
$2^q$ possible values makes $B(C)+B(S)=0$.
Thus $\Pr[B(C)=B(S)]=2^{-q}$.
A union bound over $C\in\mathcal A\setminus\{S\}$ proves the claim.
\end{proof}

\paragraph{ Entropy and total mass (Lemma~\ref{lem:entropy-mass}).}
\begin{proof}
Put $p_i=q_i/Q$, so that $\sum_i p_i=1$. Expanding the definition
of $h_N$ gives
$$
\begin{aligned}
\sum_i h_N(q_i)
&=\sum_i Qp_i\log\frac{eN}{Qp_i}\\
&=Q\log\frac{eN}{Q}\sum_i p_i
  +Q\sum_i p_i\log\frac1{p_i}\\
&=h_N(Q)+Q\mathsf H(p).
\end{aligned}
$$
Terms with $p_i=0$ contribute zero.
Since entropy is nonnegative, taking $(q_1,q_2)=(x,y)$ also gives
$h_N(x+y)\le h_N(x)+h_N(y)$ for $x,y\ge0$ and $x+y\le N$.

\end{proof}

\paragraph{ Entropy from the mean (Lemma~\ref{lem:entropy-mean}).}
\begin{proof}
Put $\mu=\mathbb ET$.
If $\mu=0$, then $T=0$ almost surely and the claim is immediate.
Assume $\mu>0$, and consider the geometric distribution
$$
g_k=\frac1{1+\mu}\left(\frac{\mu}{1+\mu}\right)^k,
\qquad k=0,1,\ldots.
$$
First suppose that $T$ has finite support, and write $p_k=\Pr[T=k]$.
Nonnegativity of relative entropy gives
$\sum_k p_k\log(p_k/g_k)\ge0$, or equivalently,
$$
\begin{aligned}
\mathsf H(T)
&\le-\sum_k p_k\log g_k\\
&=\sum_k p_k\left(\log(1+\mu)
                  +k\log\frac{1+\mu}{\mu}\right)\\
&=\log(1+\mu)+\mu\log(1+1/\mu)\\
&\le\log(1+\mu)+\log e
=\log(e(1+\mu)).
\end{aligned}
$$
The last inequality uses $\ln(1+u)\le u$ for $u\ge0$.

For general $T$, apply the same comparison with $g$ to
$T_M=\min\{T,M\}$. Since $\mathbb ET_M\le\mu$, it gives
$\mathsf H(T_M)\le\log(e(1+\mu))$.
Writing $r_M=\Pr[T\ge M]$, we have
$$
\mathsf H(T_M)
=\sum_{k=0}^{M-1}p_k\log(1/p_k)+r_M\log(1/r_M).
$$
As $M\to\infty$, the sum tends to $\mathsf H(T)$ and the last term tends to zero. Passing to the limit proves the bound.
\end{proof}

\paragraph{ Prefix-tail sums (Lemma~\ref{lem:prefix-sums}).}
\begin{proof}
Put ${ k=\lfloor z\rfloor}$. Then $G_a(z)=1$ for $0\le a\le k$,
and $G_{k+r}(z)\le2^{1-r}$ for $r\ge1$.
Consequently,
\begin{equation*}
\sum_{a\ge0}G_a(z)\le k+1+2\le { z+3},
\qquad
\sum_{a\ge0}aG_a(z)
\le\frac{k(k+1)}2+2k+4\le32(1+z)^2.
\end{equation*}
\end{proof}

%% file: main.bbl
\newcommand{\etalchar}[1]{$^{#1}$}
\begin{thebibliography}{MBLYR23}

\bibitem[Abb10]{Abb10}
Emmanuel Abbe.
\newblock Universal polar coding and sparse recovery, 2010.
\newblock arXiv:1012.0367.

\bibitem[BKW24]{BKW24}
Djamal Belazzougui, Gregory Kucherov, and Stefan Walzer.
\newblock Better space-time-robustness trade-offs for set reconciliation.
\newblock In {\em 51st International Colloquium on Automata, Languages, and
  Programming (ICALP)}, volume 297 of {\em LIPIcs}, pages 20:1--20:19, 2024.

\bibitem[CL21]{CL20}
Kuan Cheng and Xin Li.
\newblock Efficient document exchange and error correcting codes with
  asymmetric information.
\newblock In {\em Proceedings of the 2021 {ACM-SIAM} Symposium on Discrete
  Algorithms ({SODA})}, pages 2424--2443. Society for Industrial and Applied
  Mathematics, 2021.

\bibitem[CR23]{CR23}
Mahdi Cheraghchi and Jo\~{a}o Ribeiro.
\newblock Simple codes and sparse recovery with fast decoding.
\newblock {\em SIAM Journal on Discrete Mathematics}, 37(2):612--631, 2023.

\bibitem[DORS08]{DORS08}
Yevgeniy Dodis, Rafail Ostrovsky, Leonid Reyzin, and Adam~D. Smith.
\newblock Fuzzy extractors: How to generate strong keys from biometrics and
  other noisy data.
\newblock {\em SIAM Journal on Computing}, 38(1):97--139, 2008.

\bibitem[EGUV11]{EGUV11}
David Eppstein, Michael~T. Goodrich, Frank Uyeda, and George Varghese.
\newblock What's the difference? {E}fficient set reconciliation without prior
  context.
\newblock In {\em Proceedings of the ACM SIGCOMM 2011 Conference}, pages
  218--229, 2011.

\bibitem[FA25]{FA25}
Hiroki Furue and Yusuke Aikawa.
\newblock An improved {Both-May} information set decoding algorithm: Towards
  more efficient time-memory trade-offs.
\newblock In {\em Post-Quantum Cryptography (PQCrypto)}, volume 15577 of {\em
  Lecture Notes in Computer Science}, pages 104--128, 2025.

\bibitem[FLOS24]{FLOS24}
Nils Fleischhacker, Kasper~Green Larsen, Maciej Obremski, and Mark Simkin.
\newblock Invertible {Bloom} lookup tables with less memory and randomness.
\newblock In {\em 32nd Annual European Symposium on Algorithms (ESA)}, volume
  308 of {\em LIPIcs}, pages 54:1--54:17, 2024.

\bibitem[GLL{\etalchar{+}}20]{PBS20}
Long Gong, Ziheng Liu, Liang Liu, Jun Xu, Mitsunori Ogihara, and Tong Yang.
\newblock Space- and computationally-efficient set reconciliation via parity
  bitmap sketch ({PBS}).
\newblock {\em Proceedings of the VLDB Endowment}, 14(4):458--470, 2020.

\bibitem[GM11]{GM11}
Michael~T. Goodrich and Michael Mitzenmacher.
\newblock Invertible {Bloom} lookup tables.
\newblock In {\em 49th Annual Allerton Conference on Communication, Control,
  and Computing}, pages 792--799, 2011.

\bibitem[HPW23]{HPW23}
Jakob B{\ae}k~Tejs Houen, Rasmus Pagh, and Stefan Walzer.
\newblock Simple set sketching.
\newblock In {\em Symposium on Simplicity in Algorithms (SOSA)}, pages
  228--241, 2023.

\bibitem[KPW26]{KPW26}
Jonas Klausen, Rasmus Pagh, and Stefan Walzer.
\newblock Stuffed {IBLTs}: Optimal linear multiset sketches, 2026.
\newblock arXiv:2609.17487.

\bibitem[KRU11]{KRU11}
Shrinivas Kudekar, Thomas~J. Richardson, and R{\"u}diger~L. Urbanke.
\newblock Threshold saturation via spatial coupling: Why convolutional {LDPC}
  ensembles perform so well over the {BEC}.
\newblock {\em IEEE Transactions on Information Theory}, 57(2):803--834, 2011.

\bibitem[KYR25]{KYR25}
Tomer Keniagin, Eitan Yaakobi, and Ori Rottenstreich.
\newblock {CertainSync}: Rateless set reconciliation with certainty.
\newblock {\em Proceedings of the ACM on Measurement and Analysis of Computing
  Systems}, 9(2), 2025.

\bibitem[LMKY16]{LossRadar16}
Yuliang Li, Rui Miao, Changhoon Kim, and Minlan Yu.
\newblock {LossRadar}: Fast detection of lost packets in data center networks.
\newblock In {\em Proceedings of the 12th International Conference on Emerging
  Networking Experiments and Technologies (CoNEXT)}, pages 481--495, 2016.

\bibitem[MBLYR23]{MBYR23}
Avi Mizrahi, Daniella Bar-Lev, Eitan Yaakobi, and Ori Rottenstreich.
\newblock Invertible {Bloom} lookup tables with listing guarantees.
\newblock {\em Proceedings of the ACM on Measurement and Analysis of Computing
  Systems}, 7(3):61:1--61:29, 2023.

\bibitem[MTZ03]{MTZ03}
Yaron Minsky, Ari Trachtenberg, and Richard Zippel.
\newblock Set reconciliation with nearly optimal communication complexity.
\newblock {\em IEEE Transactions on Information Theory}, 49(9):2213--2218,
  2003.

\bibitem[OLB{\etalchar{+}}19]{Graphene19}
A.~Pinar Ozisik, Brian~N. Levine, George Bissias, Gavin Andresen, Darren Tapp,
  and Sunny Katkuri.
\newblock {Graphene}: Efficient interactive set reconciliation applied to
  blockchain propagation.
\newblock In {\em Proceedings of the ACM SIGCOMM 2019 Conference}, pages
  303--317, 2019.

\bibitem[Ste94]{Ste94}
Jacques Stern.
\newblock A new identification scheme based on syndrome decoding.
\newblock In {\em Advances in Cryptology---CRYPTO '93}, volume 773 of {\em
  Lecture Notes in Computer Science}, pages 13--21, 1994.

\bibitem[Str12]{Str12}
Falko Strenzke.
\newblock Fast and secure root finding for code-based cryptosystems.
\newblock In {\em Cryptology and Network Security (CANS)}, volume 7712 of {\em
  Lecture Notes in Computer Science}, pages 232--246, 2012.

\bibitem[Wal25]{Wal25}
Stefan Walzer.
\newblock Peeling close to the orientability threshold: Spatial coupling in
  hashing-based data structures.
\newblock {\em ACM Transactions on Algorithms}, 21(3):33:1--33:23, 2025.

\bibitem[XZX{\etalchar{+}}26]{XYZ26}
Rui Xu, Kangyang Zhou, Jiachen Xu, Jiarui Guo, Boyu Xian, Kaicheng Yang, Yong
  Cui, and Tong Yang.
\newblock Toward optimal time-space tradeoffs for set reconciliation, 2026.
\newblock arXiv:2609.14442, version 2.

\bibitem[YGA24]{YGA24}
Lei Yang, Yossi Gilad, and Mohammad Alizadeh.
\newblock Practical rateless set reconciliation.
\newblock In {\em Proceedings of the ACM SIGCOMM 2024 Conference}, pages
  595--612, 2024.

\bibitem[YWM{\etalchar{+}}23]{ChameleMon23}
Kaicheng Yang, Yuhan Wu, Ruijie Miao, Tong Yang, Zirui Liu, Zicang Xu, Rui Qiu,
  Yikai Zhao, Hanglong Lv, Zhigang Ji, and Gaogang Xie.
\newblock {ChameleMon}: Shifting measurement attention as network state
  changes.
\newblock In {\em Proceedings of the ACM SIGCOMM 2023 Conference}, pages
  881--903, 2023.

\end{thebibliography}
